\documentclass[a4paper,english,useAMS,usenatbib]{mnras}
\usepackage[T1]{fontenc}
\usepackage{amsmath}
\usepackage{amssymb}
\usepackage{graphicx}
\usepackage{wasysym}
\usepackage{esint}
\usepackage[authoryear]{natbib}
\usepackage[utf8]{inputenc}
\usepackage{threeparttable}
\usepackage{hyperref}
\usepackage{xspace}
\usepackage{url}
\usepackage{babel}
\usepackage{ulem}
\usepackage{newtxtext,newtxmath}
\newcommand{\arepo}{\textsc{Arepo}\xspace}
\newcommand{\Fermi}{\textit{Fermi}\xspace}
\defcitealias{2021WerhahnI}{Paper I}
\defcitealias{2021WerhahnII}{Paper II}
\defcitealias{2021WerhahnIII}{Paper III}
\newcommand{\dps}{\displaystyle}

\makeatletter

\pdfpageheight\paperheight
\pdfpagewidth\paperwidth

\title[CRs and non-thermal emission in galaxies II.]{Cosmic rays and non-thermal emission in simulated galaxies. II. $\gamma$-ray maps, spectra and the far infrared-$\gamma$-ray relation}
\author[M. Werhahn et al.]{Maria Werhahn,$^{1,2}$\thanks{E-mail:
mwerhahn@aip.de} Christoph Pfrommer,$^{1}$ Philipp Girichidis$^{1}$, Georg Winner$^{1,3}$
\\
\\
$^{1}$Leibniz-Institut f\"ur Astrophysik Potsdam (AIP), An der Sternwarte 16, 14482 Potsdam, Germany\\
$^2$Institut f\"ur Physik und Astronomie, Universit\"at Potsdam, Karl-Liebknecht-Str.\,24/25, 14476 Golm, Germany\\
$^{3}$Fakult\"at f\"ur Physik und Astronomie, Universit\"at Heidelberg, Im Neuenheimer Feld 226, 69120 Heidelberg, Germany}

\makeatother

\begin{document}
\date{Accepted 20XX . Received 20XX}

\maketitle
\pagerange{\pageref{firstpage}--\pageref{lastpage}} \pubyear{2020}

\label{firstpage}
\begin{abstract}
The $\gamma$-ray emission of star-forming (SF) galaxies is attributed to hadronic interactions of cosmic ray (CR) protons with the interstellar gas and contributions from CR electrons via bremsstrahlung and inverse Compton (IC) scattering. The relative importance of these processes in different galaxy types is still unclear. We model these processes in three-dimensional magneto-hydrodynamical (MHD) simulations of the formation of isolated galactic discs using the moving-mesh code \arepo, including dynamically coupled CR protons and adopting different CR transport models. We calculate steady-state CR spectra and also account for the emergence of secondary electrons and positrons. This allows us to produce detailed $\gamma$-ray maps, luminosities and spectra of our simulated galaxies at different evolutionary stages. Our simulations with anisotropic CR diffusion and a low CR injection efficiency at supernovae (SNe, $\zeta_{\mathrm{SN}}=0.05$) can successfully reproduce the observed far infrared (FIR)-$\gamma$-ray relation. Starburst galaxies are close to the calorimetric limit, where CR protons lose most of their energy due to hadronic interactions and hence, their $\gamma$-ray emission is dominated by neutral pion decay. However, in low SF galaxies, the increasing diffusive losses soften the CR proton spectra due to energy-dependent diffusion, and likewise steepen the pionic $\gamma$-ray spectra. In turn, IC emission hardens the total spectra and can contribute up to $\sim40$ per cent of the total luminosity in low SF galaxies. Furthermore, in order to match the observed $\gamma$-ray spectra of starburst galaxies, we require a weaker energy dependence of the CR diffusion coefficient, $D\propto{E}^{0.3}$, in comparison to Milky Way-like galaxies.
\end{abstract}


\section{Introduction}
The emergence of CRs in SF galaxies can be inferred from their non-thermal emission. Besides synchrotron emission that arises from the CR electron population and is visible in radio wavelengths, we also expect $\gamma$-ray emission in the GeV-TeV range. On the one hand, inelastic collisions of CR protons with the ambient medium are responsible for creating neutral pions that decay further into $\gamma$-ray photons. On the other hand, the CR electron population produces $\gamma$ rays either via bremsstrahlung, or by scattering off of photons from the cosmic microwave background (CMB) and other interstellar radiation fields, i.e. via IC scattering.

These processes have been predicted to occur in the nuclei of starburst galaxies, where SN rates, stellar wind powers and gas densities are expected to be large \citep{1996Voelk,1999Blom}. These conditions imply an observable $\gamma$-ray flux from SF galaxies such as NGC~253, M82 and M31 \citep{2003RomeroTorres,2008Persic,2009deCeaDelPozo,2010Rephaeli,2019McDaniel}. Indeed, these galaxies have been detected at GeV and TeV energies by several experiments, e.g., by \Fermi LAT \citep{2012AckermannGamma}, the \citet{2009VERITAS_M82} and the \citet{Abdalla2018HESS}.

The $\gamma$-ray luminosities of a sample of SF galaxies have been found to correlate with their FIR luminosities \citep{2012AckermannGamma,2016RojasBravo, 2017Linden,2020Ajello}, which traces the star formation rate (SFR). The ultra-violet (UV) light of young stellar populations gets absorbed by dust and re-emitted in the FIR, and at the same time, CRs are accelerated at remnants of supernova (SN) explosions. Hence, we expect a connection between these quantities if proton calorimetry holds \citep{1994Pohl}, similar to the well-known relation between radio and FIR luminosity \citep{Voelk1989,2010Lacki}. Calorimeter theory describes the condition that inelastic collisions in the interstellar medium (ISM) occur on short timescales so that CR protons lose most of their energy before escaping the galaxy. However, the emergence of this relation over many orders of magnitude in SFRs and its non-linearity at low SFR has not been fully understood yet.
In the case of starburst galaxies, the calorimetric assumption has been tested by modeling the spectra of individual galaxies with one-zone models \citep{2011Lacki, 2013Yoast-Hull}.
The departure from the linear relation at low SFRs has been attributed to the increasing relevance of non-radiative losses in those galaxies, e.g., adiabatic losses \citep{2017bPfrommer}, diffusion and/or advection \citep{2007Thompson,2010Strong,2011Lacki, 2012AckermannGamma, 2014Martin, 2020Kornecki}. 
However, the small number of galaxies with low SFRs detected in $\gamma$ rays makes it hard to constrain the relation in this regime. Furthermore, there might also be other contributions to the $\gamma$-ray emission of those SF galaxies, including a contribution due to a population of high-energy pulsars in the SMC \citep{2010Abdo,Lopez_2018}. Similarly, there might be pulsars and their nebulae or SN remnants contributing to the $\gamma$-ray flux of the LMC \citep{2016Ackermann_LMC}. Active galactic nuclei (AGN) can also contaminate the $\gamma$-ray flux, which has been postulated, e.g., for NGC~4945 \citep{2017Wojaczynski_NGC4945}.

In previous one-zone models, the magnetic field strength, the CR electron and proton energy densities are considered to be free parameters that are fit to reproduce observed gamma-ray and radio emission spectra \citep{2004Torres,2005Domingo-SantamariaTorres,2008Persic,2009deCeaDelPozo,2010Lacki,2011Lacki,2012Paglione,2013Yoast-Hull,2015Yoast-Hull,2016Eichmann}. Because this is an under-constrained system, a closure is assumed by either requiring that the CR proton and magnetic energy densities are in equipartition or by adopting a universal CR electron-to-proton ratio that is observed at the Solar radius in the Milky Way. More detailed one-dimensional flux-tube models \citep{2002Breitschwerdt} and two-dimensional axisymmetric models \citep{2014Martin, 2020Buckman} make use of parametrized source functions, and/or prescribed density and magnetic field distributions.

Instead, here we will simulate the $\gamma$-ray emission attributed to the star formation activity in galaxies, i.e.\ due to the emergence of CRs, by using three-dimensional MHD simulations of the formation and evolution of isolated galactic discs. Note that neither the CR proton nor the magnetic energy densities are free parameters in our approach but a product of our MHD simulations. We will be using results from kinetic plasma simulations \citep{Caprioli2014} in combination with a detailed modelling of multi-frequency emission maps and spectra of supernova remnant (SNRs) to determine the CR acceleration efficiencies \citep{Pais2018,Pais2020a,Pais2020b,2020Winner}. Equivalently, our forming galaxy drives a turbulent magnetic dynamo that amplifies the field to observed field strengths \citep{Pfrommer2021}.

In addition to only accounting for CR protons and assuming a fixed spectral index for their energy distribution \citep{2017bPfrommer}, we also account for possible changes in the spectral index due to energy dependent diffusion as well as cooling processes in this work. We revisit the production of secondary particles, such as $\gamma$ rays, secondary electrons and positrons, and also consider leptonic emission processes in the $\gamma$-ray regime from secondary leptons and primary, shock-accelerated electrons at SNe. As such, we intend to gain new insights into the physical processes governing the $\gamma$-ray emission of SF galaxies. 

This paper is the second paper of a series of three papers. It is based on the modeling of the steady-state spectra of CR electrons and protons as described in \citeauthor{2021WerhahnI} (\citeyear{2021WerhahnI}, hereafter Paper I). We recap our approach in Section~\ref{sec: methods}. We present $\gamma$-ray emission maps resulting from different production channels of our simulated galaxies in Section~\ref{sec: non-thermal gamma-ray emission from sims.}, where we analyse the FIR-$\gamma$-ray relation in Section~\ref{subsec: FIR-gamma-ray relation} and compare our $\gamma$-ray spectra to observations in Section~\ref{subsec: gamma-ray spectra}. A discussion of our results and a conclusion is presented in Section~\ref{sec: Discussion and conclusion}. In Appendix~\ref{sec: Radiation processes}, we detail the non-thermal emission processes emerging in the $\gamma$-ray regime, i.e., $\gamma$-ray emission from neutral pion decay as well as the IC and bremsstrahlung emission.

\section{Description of the Methods}  \label{sec: methods}

\subsection{Simulations}
Using the moving mesh code \arepo \citep{2010Springel,2016aPakmor}, we simulate the formation of isolated galaxies from the collapse of a gas cloud that is initially in hydrostatic equilibrium with the dark matter halo as described in \citetalias{2021WerhahnI}. The galaxies are embedded in dark matter halos that follow an NFW \citep{1997Navarro} profile with a concentration parameter of $c_{200}=12$ and masses of $M_{200}/\mathrm{M_{\odot}}=\{10^{10}, 10^{11}, 3\times 10^{11}, 10^{12} \}$. Each halo contains $10^7$ gas cells, each carrying a target mass of $155\,\mathrm{M_{\odot}}\times M_{200}/(10^{10}\,\mathrm{M_{\odot}})$ and we enforce that the mass of all cells remains within a factor of two of the target mass by explicitly refining and de-refining the mesh cells. 

The magnetic field evolution is prescribed by ideal MHD \citep{2013Pakmor}, which was shown to produce magnetic fields in cosmological simulations that match Milky Way observations \citep{Pakmor2017,Pakmor2018,Pakmor2020}. We adopt two different values for the initial seed magnetic field $B_0=\{10^{-12},10^{-10}\}\,$G that is oriented along the $x$ axis. As the gas collapses, the magnetic field experiences adiabatic compression and grows exponentially via the emergence of a turbulent small-scale dynamo \citep{Pfrommer2021}. Conservation of specific gas angular momentum causes the gas to settle in a disc out of which a galaxy forms inside out, which continues to amplify and order the disc magnetic field. 

CR protons are self-consistently included with the one-moment formalism \citep{2016cPakmor,2017aPfrommer}. We instantaneously inject CRs at the SNe with a fraction $\zeta_{\mathrm{SN}}$ of the kinetic energy of the SN explosion and model CR transport in two different ways: we either only advect them with the gas or additionally account for anisotropic CR diffusion along the magnetic field with a constant parallel diffusion coefficient of $D=10^{28}\,\mathrm{cm^2\,s^{-1}}$ or $D=3\times 10^{28}\,\mathrm{cm^2\,s^{-1}}$. In table~\ref{tab:simulations-overview}, we give an overview of the different configurations of our simulations.

The relation between the isotropic CR diffusion coefficient $D_\rmn{iso}$ and the coefficient along the magnetic field, $D$, depends on the exact magnetic field configuration. While a turbulent field implies $D_\rmn{iso} = D/3$, pure CR transport along the homogeneous magnetic field yields $D_\rmn{iso} = D$. The latter situation is realised for active CR-driven wind feedback as CRs move along open field lines from the disc into the halo and the homogeneity of the magnetic field either results from the velocity shear of the outflow or the Parker instability. CR propagation within a turbulent spiral arm of the galaxy may prefer the situation with $D_\rmn{iso} = D/3$. In fact, our adopted values for $D$ are bracketing these two cases and are consistent with the recently discovered hardening of the logarithmic momentum slope of the CR proton spectrum at low Galactocentric radii, which is interpreted as a signature of anisotropic diffusion in the Galactic magnetic field \citep{Cerri2017,Evoli2017}. Analysing AMS-02 data of unstable secondary CR nuclei that result from spallation processes in the ISM yields the residence time of CRs inside the Galaxy that constrains identical values for the diffusion coefficient \citep{Evoli2019,2020Evoli}.

We vary the energy efficiency of CR acceleration and inject CR protons directly at the location of core-collapse SN explosions with  $\zeta_{\mathrm{SN}}=5$ per cent to 10 per cent of the canonical kinetic SN energy of $10^{51}$~erg. While the larger value is a canonical value adopted in CR studies, the low efficiency is motivated by taking the acceleration efficiency of $\approx0.15$ inferred by hybrid particle-in-cell simulations of proton acceleration at quasi-parallel shocks, in which the shock normal is close to the upstream magnetic field orientation \citep{Caprioli2014} and averaging the result over the entire supernova remnant. This yields a CR proton acceleration efficiency of $\approx0.05$, independent of magnetic morphology \citep{Pais2018}. MHD simulations of expanding shell-type supernova remnants in the Sedov-Taylor phase that adopt such an efficiency for protons are able to match observational multi-frequency data \citep{Pais2020a,Pais2020b,2020Winner}. We intend to assess the effect of varying those parameters and prescriptions of CR transport on the resulting $\gamma$-ray emission in our simulated galaxies.

\begin{table}
\caption{Overview of the parameters of the different simulations with CR advection and anisotropic CR diffusion (`CR diff'): halo mass $M_{200}$, CR energy injection efficiency $\zeta_{\mathrm{SN}}$, initial magnetic field $B_0$ and parallel diffusion coefficient $D$. Note that in addition to these simulations, we also simulated all configurations with pure CR advection without accounting for diffusion (`CR adv').}
\begin{center}
\begin{tabular}{lcccc}
\hline
 $M_{200}\,[\mathrm{M_{\odot}}]$ & $\zeta_{\mathrm{SN}}$ & $B_0\,[\mathrm{G}]$& $D\,\mathrm{[cm^2/s]}$ \\
\hline
\hline
$10^{12}$         & $0.05$ & $10^{-10},\,10^{-12}$& $1\times10^{28}$\\
$3\times 10^{11}$ & $0.05$ & $10^{-10},\,10^{-12}$& $1\times10^{28}$\\
$10^{11}$         & $0.05$ & $10^{-10},\,10^{-12}$& $1\times10^{28}$ \\
$10^{10}$         & $0.05$ & $10^{-10},\,10^{-12}$& $1\times10^{28}$\\
$10^{10}$         & $0.05$ & $10^{-10}\phantom{,\,10^{-12}}$& $3\times 10^{28}$\\
\hline
$10^{12}$          & $0.10$ & $10^{-12}$& $1\times10^{28}$\\
$3\times 10^{11}$  & $0.10$ & $10^{-12}$& $1\times10^{28}$ \\
$10^{11}$          & $0.10$ & $10^{-12}$& $1\times10^{28}$ \\
$10^{10}$          & $0.10$ & $10^{-12}$& $1\times10^{28}$\\
$10^{10}$          & $0.10$ & $10^{-12}$& $3\times 10^{28}$\\

\hline
\label{tab:simulations-overview}
\end{tabular}
\end{center}
\end{table}

\subsection{Steady-state spectra}
We follow the approach described in \citetalias{2021WerhahnI}, where we calculate steady-state spectra $f(E)=\mathrm{d}N/(\mathrm{d}E\,\mathrm{d}V)$ in each cell of our simulations by solving the diffusion-loss equation for CR protons, primary and secondary electrons, respectively. Following e.g., \citet{1964ocr..book.....G} and \citet{2004Torres}, this reads
\begin{align}
\frac{\mathrm{}f(E)}{\tau_{\mathrm{esc}}}-\frac{\mathrm{d}}{\mathrm{d}E}\left[f(E)b(E)\right]=q(E),
\label{eq:diff-loss-equ}
\end{align}
where $E$ denotes the CR energy.
The injection spectrum $q(E)= q[p(E)] \rmn{d}p/\rmn{d}E$ is assumed to be a power law in momentum for CR protons as well as primary electrons, with the same spectral index $\alpha_{\mathrm{inj}}=2.2$ \citep{2013LackiThompson}. Furthermore, we assume an exponential cutoff in the source functions given by
\begin{align}
q_{i}(p_{i})\mathrm{d}p_{i} = C_{i} p_{i}^{-\alpha_{\mathrm{inj}}} \exp[-(p_i/p_{\mathrm{cut},i})^{n}]\mathrm{d}p_{i},
\label{eq: source fct. Q(p)}
\end{align}
where $i=\mathrm{e,p}$ denotes the CR species and $n=1$ for protons and $n=2$ for electrons \citep{2007Zirakashvili,2010Blasi}. The cutoff momenta are for protons $p_{\mathrm{cut,p}}=1\,\mathrm{PeV}/m_{\mathrm{p}}c^2$  \citep{1990Gaisser} and electrons $p_{\mathrm{cut,e}}=20\,\mathrm{TeV}/m_{\mathrm{e}}c^2$ \citep{2012Vink}.

For CR protons, we consider energy losses, $b(E)=-\mathrm{d}E/\mathrm{d}t$, due to hadronic losses and Coulomb interactions. After also accounting for CR escape due to advection and diffusion we re-normalise the steady-state spectra to match the CR energy density in each cell. The escape losses include losses due to advection and diffusion, i.e., 
\begin{align}
\tau_{\mathrm{esc}}=\frac{1}{\tau_{\mathrm{adv}}^{-1} + \tau_{\mathrm{diff}}^{-1}}.
\label{eq:tau_esc}
\end{align}
The diffusion timescale is estimated using an estimate for the diffusion length in each cell, $L_{\mathrm{CR}}=\varepsilon_{\mathrm{CR}}/\left|\nabla\varepsilon_{\mathrm{CR}}\right|$, via 
\begin{align}
\tau_{\mathrm{diff}}=\frac{L_{\mathrm{CR}}^{2}}{D}.
\label{eq:tau_diff}
\end{align}
Furthermore, we assume an energy dependent diffusion coefficient $D=D_0 (E/E_0)^{\delta}$, where we use $D_0=10^{28}$ and $3\times10^{28}\,\mathrm{cm^2~s}^{-1}$ and $E_0=3\,\mathrm{GeV}$. The energy dependence of our diffusion coefficient is assumed to be $\delta = 0.5$  in our fiducial model, which has been obtained by fitting observed beryllium isotope ratios \citep{2020Evoli}. We also study how modifying this parameter impacts the resulting $\gamma$-ray emission. We calculate the timescale of advection via
\begin{align}
\tau_{\mathrm{adv}}=\frac{L_{\mathrm{CR}}}{\varv_{z}},
\label{eq:tau_adv}
\end{align}
where we only take into account cell velocities away from the disc in $z$-direction. This is justified because mass fluxes in and out of cells in the azimuthal direction nearly compensate each other (see figure~6 of \citetalias{2021WerhahnI}). Hence, in the cell-based approximation, only the advection perpendicular to and away from the disc is relevant in order to estimate the advection losses. Note that radial CR transport via advection and anisotropic diffusion is also strongly suppressed because of the largely toroidal magnetic field configuration in the disc \citep{2013Pakmor,2016bPakmor_winds} and the dominant kinetic energy density associated with the toroidal velocity component \citep{Pfrommer2021}. Any residual CR fluxes not explicitly modeled in our steady-state approach need to be simulated by evolving the CR electron and proton spectra in our MHD simulations \citep{2019Winner,2020Winner,2020MNRAS.491..993G}.

In addition to escape losses, CR electrons can also lose energy due to the emission of radiation. Hence, their energy loss terms $b(E)$ additionally include synchrotron, IC and bremsstrahlung losses. The synchrotron and IC losses have the same dependence on energy, i.e.\ in the relativistic regime we obtain $b_\rmn{syn}\propto B^2 E^2$ and $b_\rmn{IC}\propto B_{\rmn{ph}}^2 E^2$ (where $B$ and $B_{\rmn{ph}}=\sqrt{8\upi \varepsilon_\rmn{ph}}$ are the strengths of the magnetic field and equivalent magnetic field of a photon distribution with an energy density $\varepsilon_\rmn{ph}$, respectively), whereas bremsstrahlung losses scale as $b_\rmn{brems}\propto n_\rmn{p} E \ln(2E)$ (where $n_\rmn{p}$ is the proton number density, see \citetalias{2021WerhahnI} for details). Furthermore, the primary electron population is tied to the protons by means of an injected ratio of electrons to protons, i.e.\ $K_{\mathrm{ep}}^{\mathrm{inj}}=0.02$, which is chosen so that it reproduces the observed value in the Milky Way  at 10 GeV, when averaging over the CR spectra around the solar galacto-centric radius in a snapshot that resembles the Milky Way in terms of halo mass and SFR (see \citetalias{2021WerhahnI} for a more detailed discussion). 

Inelastic collisions of CR protons with the ISM generate a secondary population of CR electrons and positrons. We calculate their production spectra using different approaches. For small kinetic proton energies, i.e., $T_{\mathrm{p}}<10\,\mathrm{GeV}$, we use the model by \citet{2018Yang} for the normalised pion energy distribution and combine it with our own parametrization of the total cross section for $\pi^{\pm}$ production that is provided in \cite{2021WerhahnI}. At high energies, we adopt the description by \citet{2006PhRvD..74c4018K} for $T_{\mathrm{p}}>100\mathrm{\,GeV}$ and perform a cubic spline interpolation in the energy range in between.

In \citetalias{2021WerhahnI}, we study the validity of the cell-based steady-state assumption. To this end, we calculate the characteristic timescale of the change in total energy density of CRs in our simulations, $\tau_{\mathrm{CR}}=\varepsilon_{\mathrm{CR}}/ \dot{\varepsilon}_{\mathrm{CR}}$. In order to maintain a steady state, we require that all cooling or escape processes in the diffusion-loss equation are faster than that timescale, i.e., $\tau_{\mathrm{all}} \lesssim \tau_{\mathrm{CR}}$. Here, $\tau_{\mathrm{all}}$ is the combined rate of all relevant cooling  and diffusion processes at a given energy, i.e., $\tau_{\mathrm{all}}^{-1}= \tau_{\mathrm{cool}}^{-1}+\tau_{\mathrm{diff}}^{-1}$.\footnote{Except for fast outflows, the advection timescale is larger than the diffusion timescale throughout the galaxy as is shown in figure~7 of \citet{2021WerhahnI}. This justifies our neglect of advection in $\tau_{\mathrm{all}}$.} In figure~9 of  \citetalias{2021WerhahnI}, we find that the steady-state approximation breaks down in regions of low gas density, in regions surrounding SNRs that host freshly injected CRs, and in outflows: these are all situation that lead to fast changes in the CR energy density, which disturb the steady-state configuration and would require to dynamically evolve the spectral CR proton and electron distributions \citep{2019Winner,2020MNRAS.491..993G,2020arXiv200906941O}. However and most importantly, weighting each Voronoi cell by the non-thermal radio synchrotron or hadronic gamma-ray emission reshapes the distribution in such a way, that the absolute majority of non-thermally emitting cells obey the steady-state condition: $\tau_{\mathrm{all}} \lesssim \tau_{\mathrm{CR}}$. This implies that the steady-state assumption is well justified in regions that dominate the non-thermal emission.

\subsection{Non-thermal emission processes in the gamma-ray regime}
\label{sec: NT emission}

We consider the following processes contributing to the emission of $\gamma$-rays from CRs:
\begin{align}
j_E=E\frac{\mathrm{d}N_{\gamma}}{\mathrm{d}E \mathrm{d}V \mathrm{d}t}=j_{E,\pi^0} + j_{E,\mathrm{IC}}+ j_{E,\mathrm{brems}},
\end{align}
where $N_{\gamma}$ denotes the number of produced photons with energy $E$ per unit energy, volume and time.
The first term arises from hadronic interactions of CR protons with the ISM, that produce neutral pions, that in turn decay further into two $\gamma$-ray photons. To calculate the resulting emissivity $j_{E,\pi^0}$, we adopt the parametrizations given by \citet{2018Yang} for small proton kinetic energies $T_\rmn{p}<10\,\mathrm{GeV}$ and the model by \citet{2014PhRvD..90l3014K} at larger energies.
Furthermore, the primary and secondary CR electron populations give rise to two additional emission processes. In the presence of an ambient radiation field, low-energy photons can be up-scattered by CR electrons via IC scattering to gamma-ray energies. We calculate the emitted IC emissivity $j_{E,\mathrm{IC}}$ including the Klein-Nishina formalism, following \citet{1968PhRv..167.1159J} and \citet{1970BlumenthalGould}. For the incident radiation field scattering off of CR electrons, we take into account the radiation from the CMB and from stars. We assume that the latter is dominated by FIR emission, which is absorbed and re-emitted UV radiation from young massive stars that are enshrouded by dusty environments. The FIR emission is characterised by a black body temperature of $T=20\,\mathrm{K}$ (see Appendix~\ref{subsec:IC-emission} for details).
Additionally, the acceleration of CR electrons in the field of charged nuclei causes the emission of bremsstrahlung, denoted by $j_{E,\mathrm{brems}}$, that we describe in the Born approximation for non-screened ions and for highly relativistic electrons \citep{1970BlumenthalGould}. We detail our calculation of these radiation processes in Appendix~\ref{sec: Radiation processes} and compare it to two other models from the literature in Appendix~\ref{sec: comparison to other models}.

\section{Non-thermal gamma-ray emission from simulated galaxies}  \label{sec: non-thermal gamma-ray emission from sims.}

\begin{figure*}
\begin{centering}
\includegraphics[scale=1.]{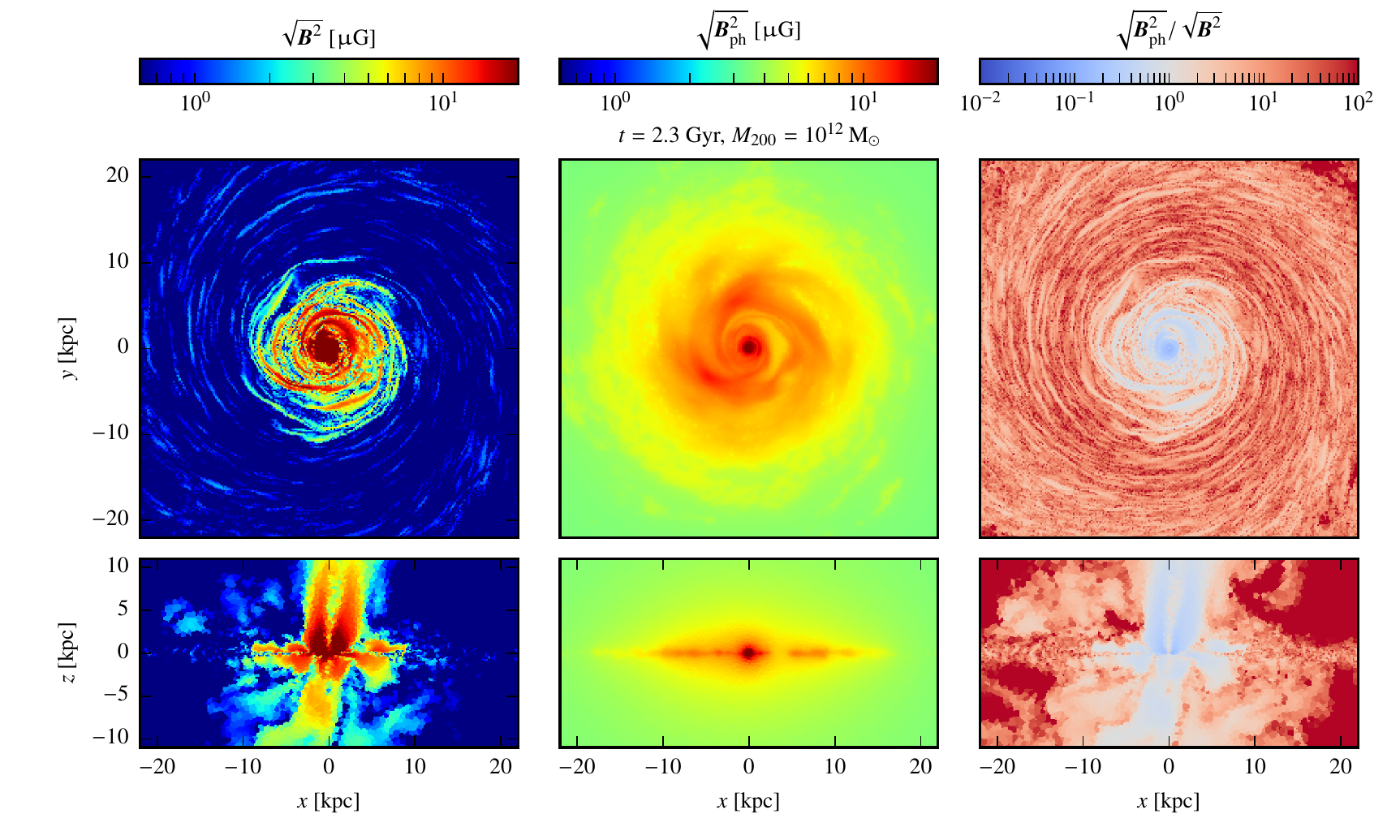}
\par\end{centering}
\begin{centering}
\includegraphics[scale=1.]{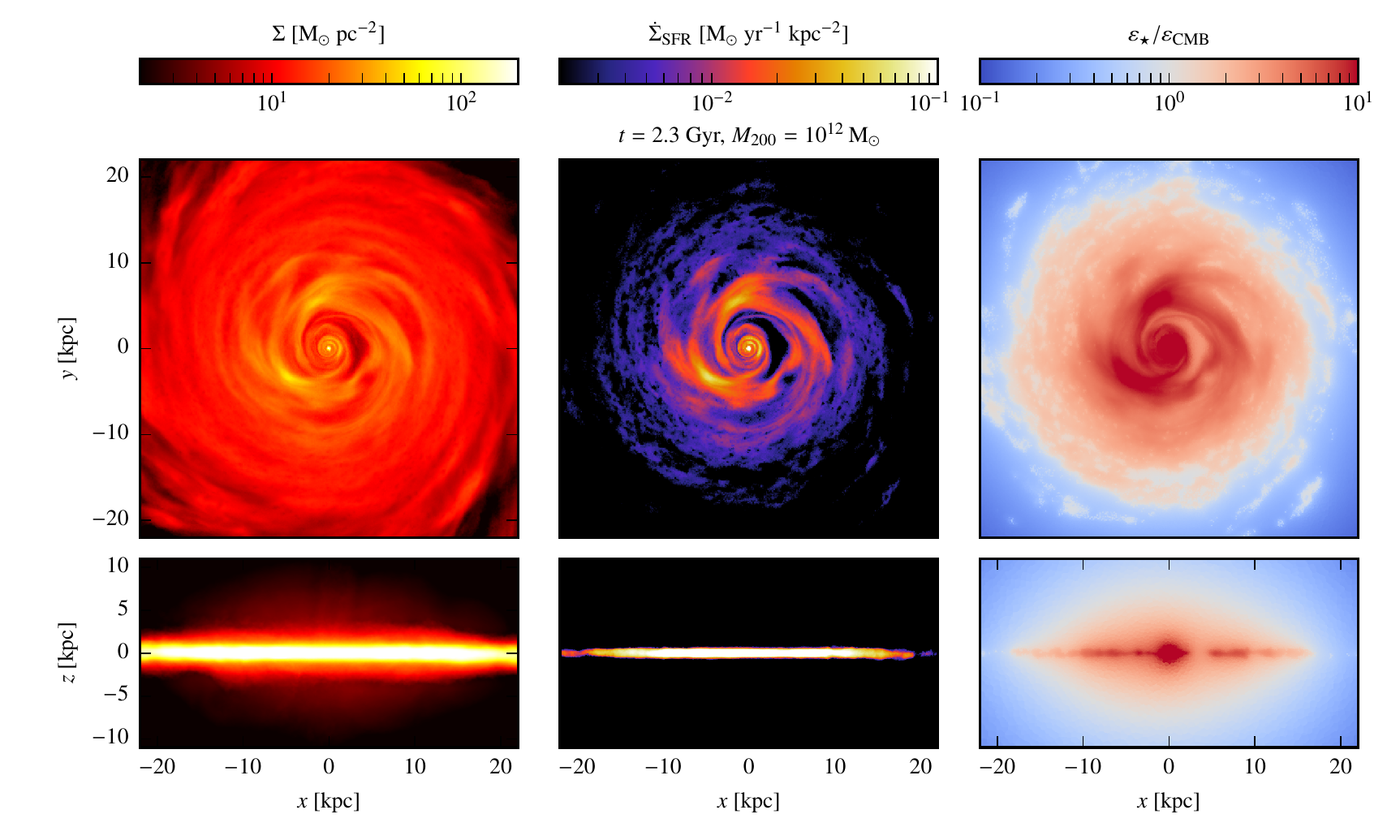}
\par\end{centering}
\caption{Face-on and edge-on maps of slices of the magnetic field, the equivalent magnetic field of the photon energy density, i.e. $B_{\rmn{ph}}=\sqrt{8\upi \varepsilon_\rmn{ph}}$, and their ratio (upper panels, from left to right). The lower panels show (from left to right) the projected gas surface density, the star-formation rate density and the ratio of the photon energy density from the stellar radiation field to the CMB.} 
\label{fig:Maps-properties}
\end{figure*}

We now apply the prescriptions for the three main radiation processes in the $\gamma$-ray regime (Section~\ref{sec: NT emission}) to our simulated galaxies. To calculate the $\gamma$-ray emission from neutral pion decay, we use the steady-state spectra of CR protons $f_{\mathrm{p}}$ in each cell and solve the integral of Eq.~(\ref{eq:production of gamma-rays}). Similarly, the cell-based leptonic steady-state spectra, $f_{\mathrm{e}}$, consisting of primary and secondary contributions, enable us to determine the resulting IC and bremsstrahlung emission from Eq.~(\ref{eq: IC emission}) and (\ref{eq:j_brems}). Here, we make use of the physical quantities in each cell, i.e.\ the gas density (for hadronic interactions and bremsstrahlung emission) and our approximation for the photon radiation field (for the IC emission). Those properties are shown in Fig.~\ref{fig:Maps-properties} for a snapshot of a simulation with a halo mass of $M_{200}=10^{12}\,\mathrm{M_{\odot}}$ after $t=2.3\,\mathrm{Gyr}$ that includes anisotropic diffusion of CR protons. This snapshot exhibits a SFR, $\gamma$-ray and radio luminosity similar to M82 (see Table \ref{Table-Galaxies}). In addition, the halo mass of $M_{200}=10^{12}\,\mathrm{M_{\odot}}$ seems to be a reasonable assumption for M82. The stellar mass can be derived from the K-band apparent magnitude, which we take from the Two Micron All Sky Survey \citep[2MASS;][]{2006Skrutskie} Extended Source Catalog (XSC) to be $m_K=4.665$. Adopting a distance of  3.7~Mpc yields a total magnitude of $M_K=-23.176$, from which we estimate the stellar mass of M82 of $\log_{10}M_{\star}\approx 10.66$, using the relation obtained by \citet{2013Cappellari}. Adopting the \citet{2010Moster} relation between stellar and halo mass of galaxies yields a halo mass of $1.6\times 10^{12}\,\mathrm{M_{\odot}}$.

\begin{table*}
 \caption{Overview of the individual observed galaxies, whose spectra are compared to simulated galaxies in Fig.~\ref{fig:gamma-ray-spectra-data}. The SFRs have been calculated by \citet{2020Kornecki}, using far UV \citep{2007GilDePaz,2012Cortese} and IRAS $25\,\mathrm{\umu m}$ data \citep{2003Sanders}. The observed $\gamma$-ray and FIR-luminosities are taken from \citet{2020Ajello}. The FIR luminosities from our simulated galaxies are inferred using the \citet{1998Kennicutt} relation. }
 \label{Table-Galaxies}
 \begin{threeparttable}[t]
 \begin{tabular}{lcccc}
  \hline
  Galaxy & SFR (obs./sim.) & $L_{\gamma}$ (obs./sim.) & $L_{\mathrm{FIR}}$ (obs./sim.) & Simulation\\
  & $[\mathrm{M}_{\odot}~\mathrm{yr}^{-1}]$&$[\mathrm{erg~s}^{-1}]$&$[\mathrm{L}_{\odot}]$ & $M_{200}\,[\mathrm{M_{\odot}}],~t\,[\mathrm{Gyr}]$ \\
  \hline
    \hline
   NGC 2146 & $14.0\pm 0.5$\tnote{1} &  $8.86\times 10^{40} $ & $1.17\times 10^{11}$ & - \\
                      & $25.520$ & $1.04\times 10^{41}$ 
                      & $1.90\times 10^{11}$ & $M_{200}=10^{12}$, $t=0.7$ \\
   M82\tnote{2} &  $10.4\pm 1.6$\tnote{1} &  $1.85\times 10^{40} $& $5.89\times 10^{10}$& - \\
                      &  $6.457$  & $2.62\times 10^{40}$  
                      & $4.81 \times 10^{10}$ & $M_{200}=10^{12}$, $t=2.3$ \\
   NGC 253  &   $5.03 \pm 0.76$\tnote{1} &  $1.16\times 10^{40} $ & $2.75\times 10^{10} $& - \\
                     &   $4.110$        &  $1.25\times 10^{40} $ 
                     & $3.06 \times 10^{10}$& $M_{200}=3\times 10^{11}$, $t=1.1$\\
   SMC         & $0.027 \pm 0.003$\tnote{1}  & $1.38\times 10^{37} $ & $7.24 \times 10^7$& -\\
                     & $0.011$                 & $1.34\times 10^{37} $ & $8.10 \times 10^7$& $M_{200}=10^{10}$, $t=2.3$\\ 
  \hline
 \end{tabular}
 \begin{tablenotes}
      \item[1] \citet{2020Kornecki}.
      \item[2] See corresponding maps in Fig.~\ref{fig:Maps-properties} and \ref{fig:Maps-gamma-ray}.
   \end{tablenotes}
 \end{threeparttable}
\end{table*}

The map of the magnetic field in Fig.~\ref{fig:Maps-properties} is shown after 2.3~Gyr of evolution and still traces the outflow that has already been launched at $\sim$1~Gyr. At that time, the increasing CR pressure gradient sourced by the ongoing injection of CR protons at SN remnants within the disc launched a galactic outflow that eventually forced the magnetic field lines to open up. After 2.3~Gyrs, this feature is still imprinted in the morphology of the magnetic field, whereas the CR outflow has already dissolved due to the decreasing SFR and hence the decreasing CR injection over time.

Contrary to the peculiar morphology of the magnetic field, the galactic photon field that is composed of interstellar radiation and CMB as characterised by an equivalent magnetic field is rather homogeneously distributed, and persists well above and below the disc, where we also have CR electrons.\footnote{For reference, the equivalent magnetic field of the CMB is 3.24~$(1+z)^2\,\umu$G at cosmological redshift $z$, which at the present epoch corresponds to the light green colour in the top middle panel of Fig.~\ref{fig:Maps-properties}.} To order of magnitude, we can estimate the equivalent magnetic field of the stellar radiation energy density, $B_{\star}$. Using the SFR of 6.48 $\rmn{M_\odot~yr}^{-1}$ of the $10^{12}~\rmn{M}_\odot$ model at 2.3 Gyr (shown in Fig.~\ref{fig:Maps-properties}), this corresponds to a FIR luminosity of $L_\rmn{FIR} = 4.8\times 10^{10} \rmn{L}_\odot$ (see Eq.~\ref{eq: radiation field}). Approximating the star formation as a uniform thin disk of radius $r$, we obtain
\begin{align}
\frac{B_{\star}^2}{8\upi} = \frac{F_\mathrm{FIR}}{c} = \frac{L_\mathrm{FIR}}{2\upi r^2c}
\end{align}
and hence
\begin{align}
B_\star \approx  5 \left(\frac{L_{\mathrm{FIR}}}{4.8\times 10^{10} \mathrm{L}_{\odot}}\right)^{1/2} \left(\frac{r}{10~\mathrm{kpc}}\right)^{-1} \,\umu\mathrm{G}.
\end{align}
Adding the CMB equivalent magnetic field, we obtain the equivalent magnetic field of the galactic photon field, $B_\mathrm{ph}=\sqrt{B_\star^2+B_\mathrm{CMB}^2}$, which ranges from $B_\mathrm{ph}\approx 6\,\mathrm{\umu G}$ at around 10 kpc to $B_\mathrm{ph}\approx 4\,\mathrm{\umu G}$ at around 20 kpc (see top-middle panel of Fig.~\ref{fig:Maps-properties}). The bottom right-hand panel of  Fig.~\ref{fig:Maps-properties} shows the stellar-to-CMB energy density ratio: the stellar radiation energy density dominates over that of the CMB inside a galactocentric radius of $r\lesssim20$~kpc, which reinforces the need to reliably model this contribution as its morphology is imprinted into the IC gamma-ray emission.

The top panel on the right-hand side of Fig.~\ref{fig:Maps-properties} shows the ratio $B_\mathrm{ph}/B$ that indicates the relative importance of IC and synchrotron processes: while the central regions at $r\lesssim5$~kpc has a dominant magnetic field with $B\gtrsim B_\mathrm{ph}$ so that synchrotron emission dominates over IC radiation, the situation is reversed at larger radii. Hence, we also expect the IC losses to be dominant over synchrotron losses at these larger radii $r\gtrsim5$~kpc in the disc and outside the outflows in which synchrotron effects dominate because of the strong magnetic field.

In addition, in the bottom panels of Fig.~\ref{fig:Maps-properties} we show the gas column density and the SFR that closely traces the former. This is a consequence of our ISM, which is modelled with an effective equation of state \citep{2003SpringelHernquist}. In this model, the star-forming gas is treated as a two phase medium in which star formation occurs in thermally unstable dense gas above a critical threshold density of $n_\mathrm{th} = 0.13\, \rmn{cm}^{-3}$ in a stochastic manner with a probability that scales exponentially with time.

\begin{figure*}
\begin{centering}
\includegraphics[scale=1.]{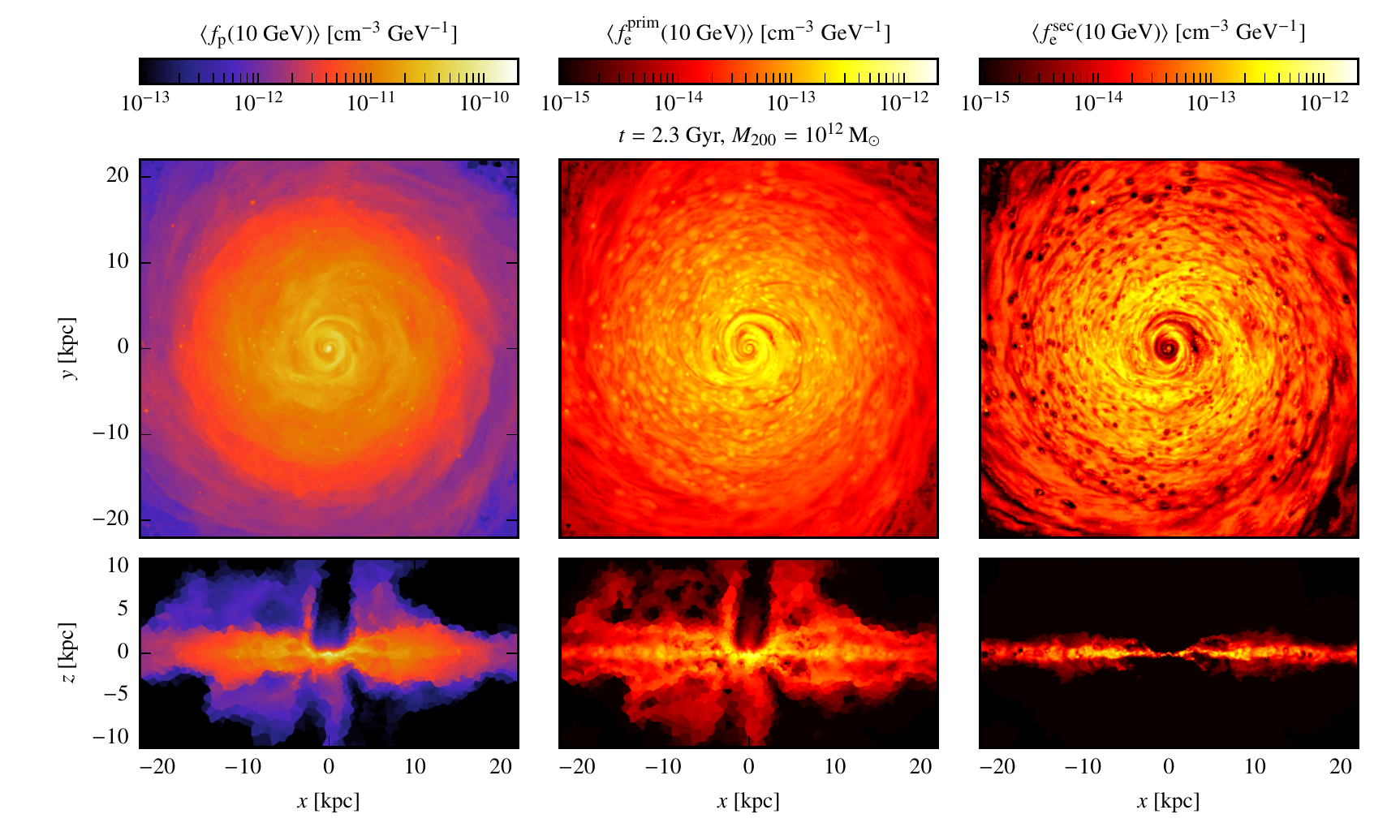}
\par\end{centering}
\caption{Face-on and edge-on maps of the spectral density of CR protons and primary and secondary CR electrons (at 10 GeV), averaged over a slice with thickness $0.3\,\mathrm{pc}$, for the same halo as in Fig.~\ref{fig:Maps-properties}.} 
\label{fig:Maps-properties_2}
\end{figure*}

Figure~\ref{fig:Maps-properties_2} shows the spectral density of CR protons, primary and secondary electrons at 10~GeV. Each of these face-on maps reveal the close morphological correspondence of dense star-forming regions (in form of spiral structures) and the produced primary electrons and protons, see Fig.~\ref{fig:Maps-properties}. Interestingly, the secondary CR electron maps exhibits low-density cavities that correspond to locations of SNR bubbles that form as a consequence of young stellar populations and which have freshly injected CR protons. The edge-on views of the different CR population reveal striking differences: while primary CR electron and proton maps are puffed up and show a CR-driven galactic wind,\footnote{The morphology of primary CR electrons become progressively more uncertain in galactic outflows (at large distances from sources) where the steady-state condition is not fulfilled and because the source function does not anymore represent generic CR sources such as SNRs but net gains due to advection. We postpone simulations that explicitly follow the electron distribution in space and time \citep{2019Winner,2020Winner} to future work.} the secondary CR electron map is tightly constrained to the dense ISM because its source function scales as $f_\rmn{e}^\rmn{sec}\propto n_\rmn{N} f_\rmn{p}$, where $n_\rmn{N}$ denotes the number density of target nucleons in the ISM, see Appendix~\ref{subsec:Gamma-ray-emission-from neutral pion decay}.

\begin{figure*}
\begin{centering}
\includegraphics[scale=1.]{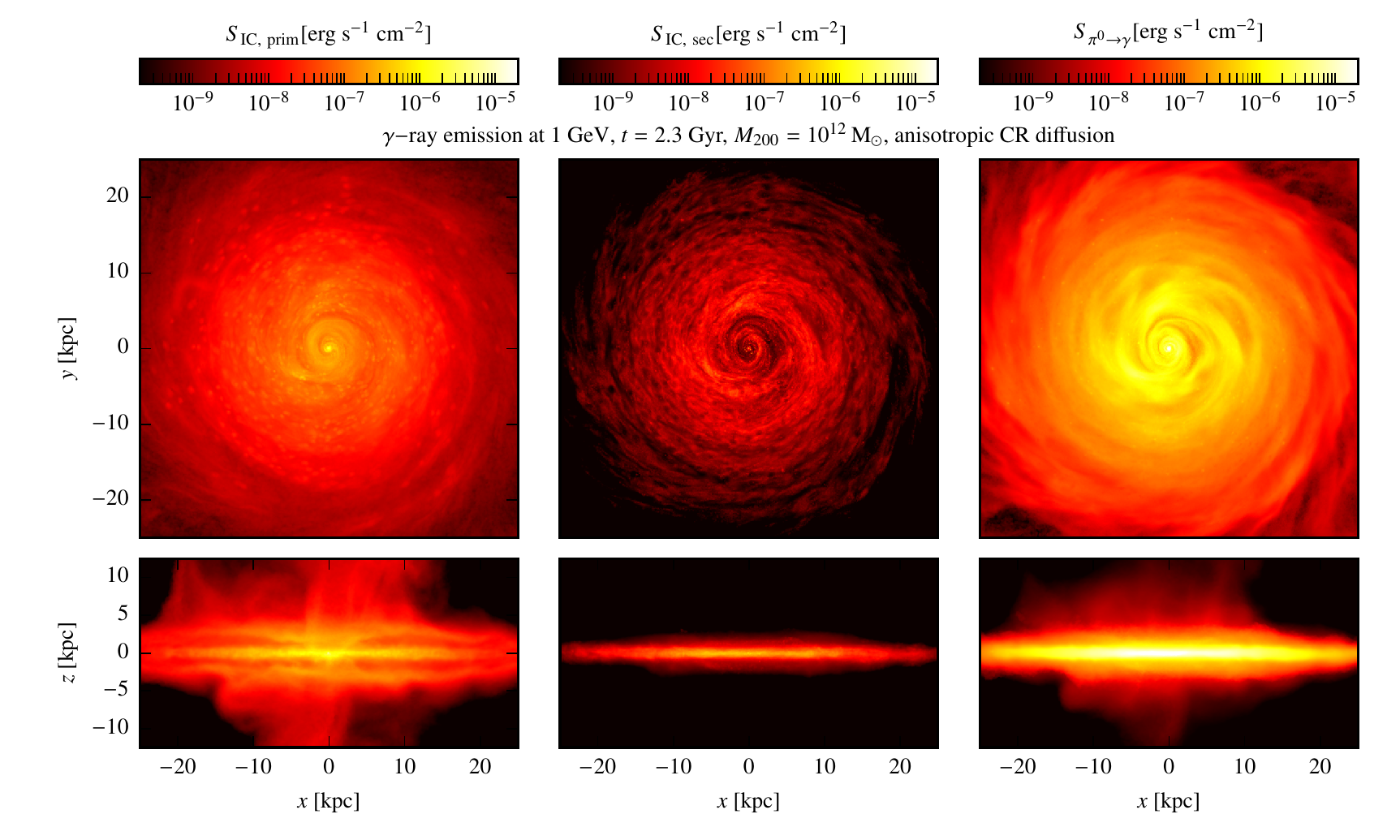}
\par\end{centering}
\begin{centering}
\includegraphics[scale=1.]{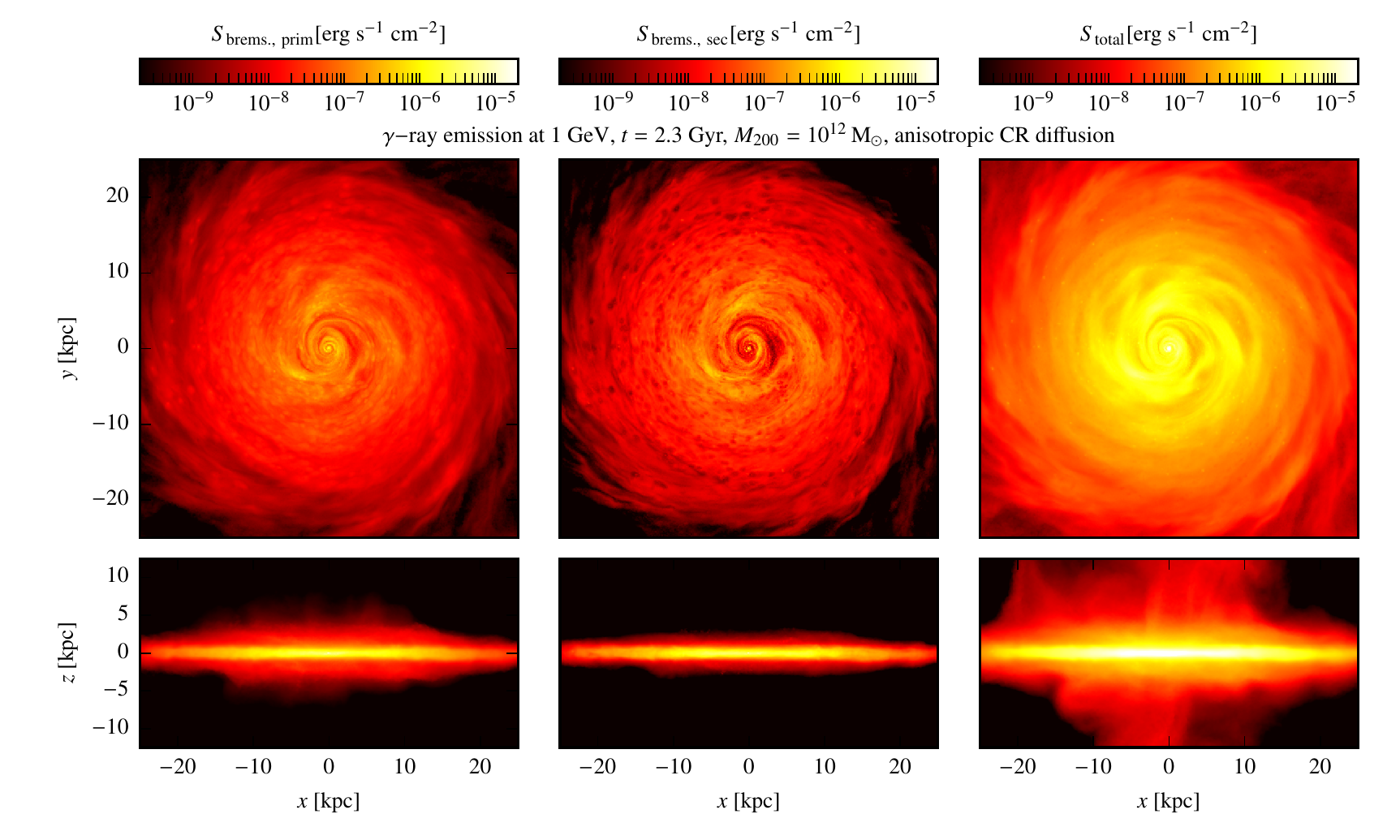}
\par\end{centering}
\caption{Projected maps of the $\gamma$-ray emission at 1\,GeV from the same snapshot as shown in Fig.~\ref{fig:Maps-properties}. From left to right, we show the different contributions, i.e. the primary and secondary IC emission, the emission from neutral pion decay in the upper panels and the primary and secondary bremsstrahlung as well as the total $\gamma$-ray emission in the lower panels.} 
\label{fig:Maps-gamma-ray}
\end{figure*}

The resulting maps of the different radiation processes at an energy of $1\, \mathrm{GeV}$ are shown in Fig.~\ref{fig:Maps-gamma-ray} for the different components. To this end, we project the emissivities along the line of sight to obtain face-on and edge-on views of the simulated galaxy. The IC emission is dominated by the emission resulting from primary CR electrons, that also reside outside the disc, where the photon radiation field acting as incoming photons for IC scattering is still strong. In contrast, secondary electrons can only be effectively produced within the disc, where the gas density is high (see column density panel in Fig.~\ref{fig:Maps-properties}), which thus also confines their IC emission close to the mid-plane. Similarly, the $\gamma$-ray emission from neutral pion decay is strongest within the disc, where the gas and CR proton densities are both high.
In contrast to the IC emission, the bremsstrahlung emission arising from primary and secondary electrons show similar strengths of their projected emissivities and in both cases, the emission approximately mimics the morphology of the gas density. This results from a similar occurrence of primary and secondary electrons in the disc \citepalias[as discussed in][]{2021WerhahnI} at 10~GeV and hence, their emitted bremsstrahlung is found to be comparably strong.
Finally, the total emission at $1\,\mathrm{GeV}$, that is shown in the lower right panel, is dominated by hadronic emission within the disc and the central regions of the galaxy, whereas IC emission from primary electrons is the main contributor to the total emission above and below the mid-plane.

\subsection{The FIR-$\gamma$-ray-Relation} \label{subsec: FIR-gamma-ray relation}

\subsubsection{Observations}
As CRs are injected at SNRs, we expect a connection between the SFR and the $\gamma$-ray emission of SF galaxies. In fact, this as been found by \citet{2012AckermannGamma} and \citet{2016RojasBravo}. Recently, \citet{2020Ajello} revisited the $\gamma$-ray luminosity observed with \Fermi LAT of 11 bona-fide $\gamma$-ray emitting galaxies. We plot their observations (black points) in the upper panels of Fig.~\ref{fig:FIR-gamma-ray,fraction_Lgamma}, converting their FIR-luminosites to SFRs using \citet{1998Kennicutt}, except for the SMC, LMC and M33. It is well known that for those low-SFR galaxies, the \citet{1998Kennicutt} conversion from FIR-luminosity to SFR does not hold anymore. This has been particularly pointed out by \citet{2020Kornecki}, who discuss this effect in more detail.
Hence, in the upper panels of Fig.~\ref{fig:FIR-gamma-ray,fraction_Lgamma} we plot the SFR (black points) and FIR luminosities (grey points) separately for the SMC, LMC and M33, where the deviations from the simple FIR-to-SFR conversion are expected to be the largest. 
The SFR of M33 was recently investigated by \citet{Thirlwall2020}, who obtained a value of $0.28^{+0.02}_{-0.01}\,\mathrm{M_{\odot}\,yr^{-1}}$. 
Reconstructing the star formation history of the LMC, \citet{2009HarrisZaritsky} deduce a lower limit for its current SFR of $0.2\,\mathrm{M_{\odot}\,yr^{-1}}$. Consistently, \citet{2020Kornecki} infer a SFR of $0.20\pm 0.03\,\mathrm{M_{\odot}\,yr^{-1}}$ from its observed far UV \citep{2012Cortese} and IRAS $25\,\umu \mathrm{m}$ \citep{2003Sanders} fluxes.
In the same way, they obtain for the SMC a SFR of $0.027\pm 0.003\,\mathrm{M_{\odot}\,yr^{-1}}$. 

It is important to keep in mind that the existing sample of SF galaxies that have been observed in $\gamma$ rays is quite small, especially towards low SFRs, where the FIR-luminosity to SFR-conversion starts to break down. Hence, more observations of these galaxies with low SFRs are needed in order to better constrain the relation. In addition, there are still some galaxies included in the analysis of the FIR-$\gamma$-ray relation, that are suspected of hosting an AGN, that could give a significant contribution to the $\gamma$-ray luminosity, biasing the interpretation that the observed $\gamma$-ray emission is solely arising from SF processes in these galaxies.

\begin{figure*}
\includegraphics[]
{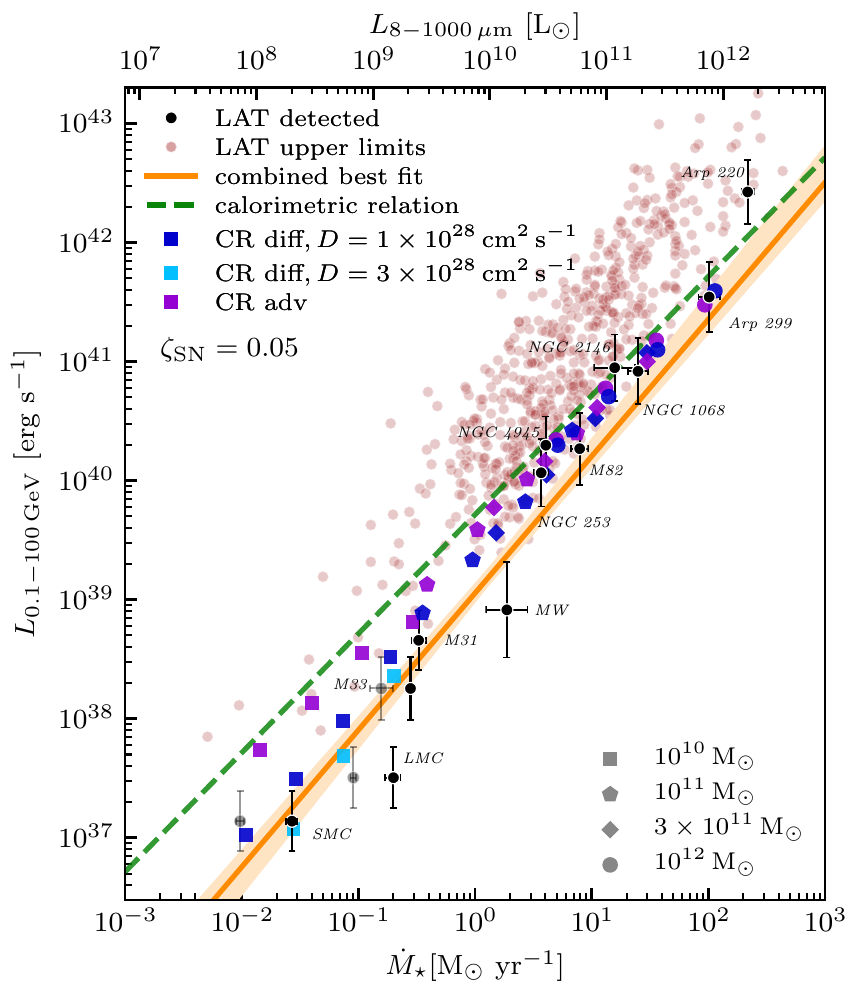}\includegraphics[]
{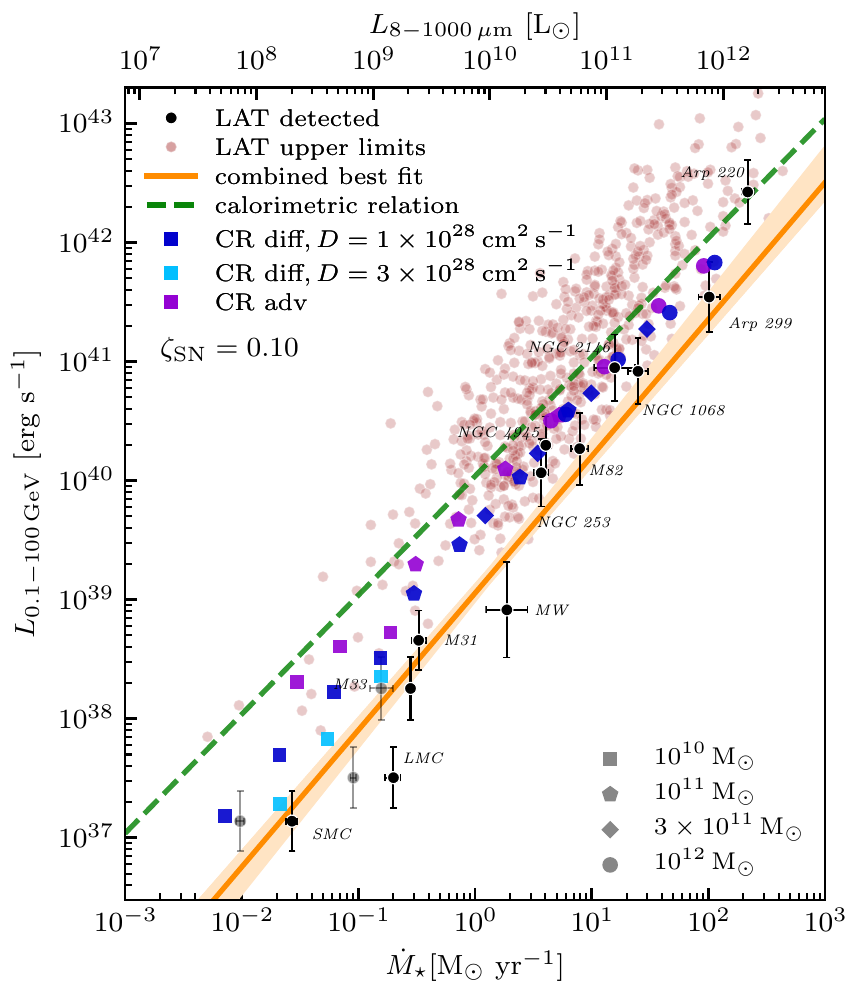}
\includegraphics[]
{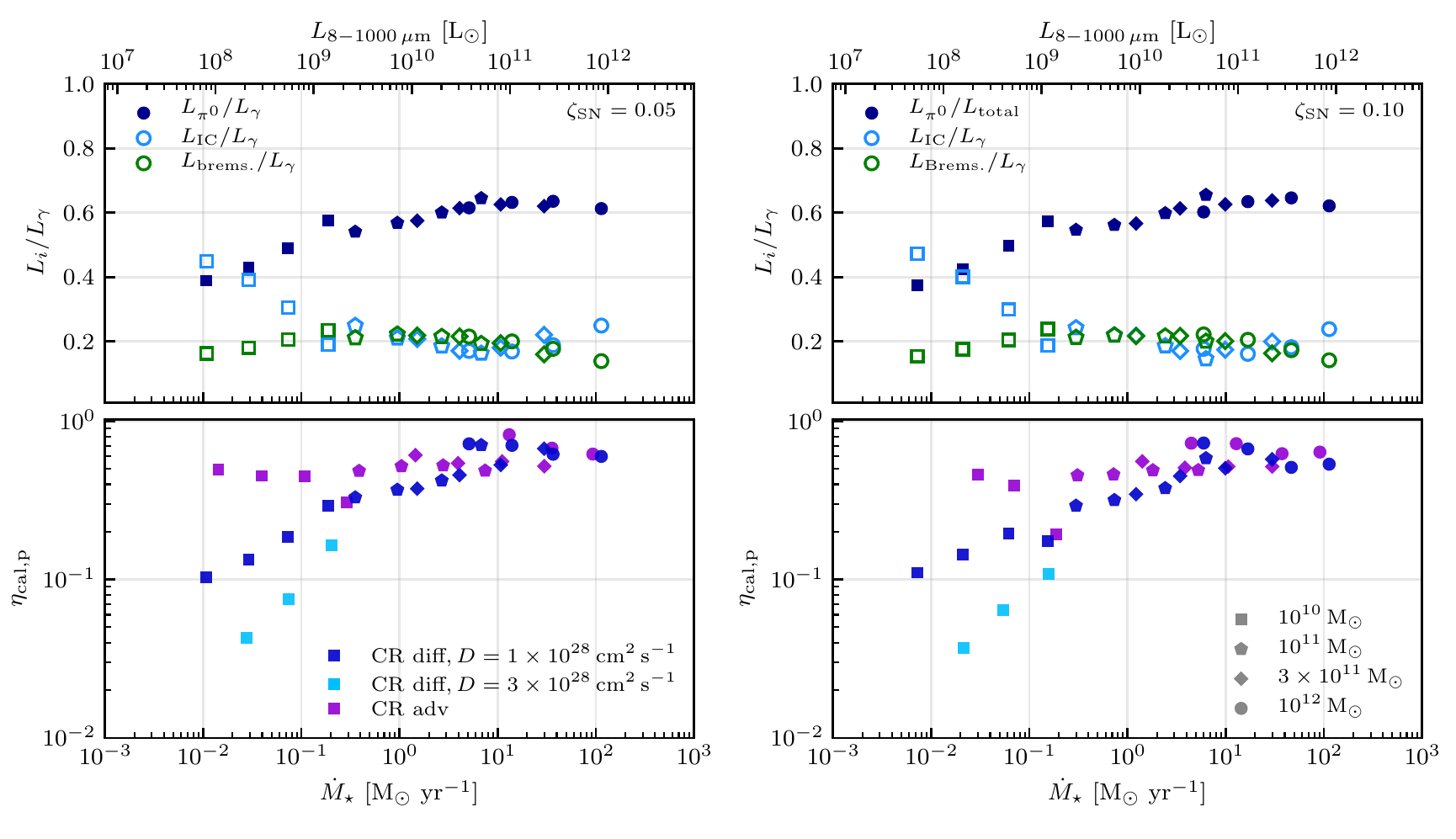}
\caption{Upper panels: The FIR-$\gamma$-ray relation for our simulated galaxies, with $\zeta_{\mathrm{SN}}=0.05$ (left panel) and  $\zeta_{\mathrm{SN}}=0.10$ (right panel). We contrast a model that only accounts for CR advection (`CR adv', purple) to the model that additionally includes anisotropic diffusion with different diffusion coefficients (`CR diff', dark and light blue). The calorimetric relation (dashed green line) is normalised to $L_{\gamma}/\eta_\rmn{cal,p}$ of the simulation with the highest SFR. Additionally, we show the FIR- and $\gamma$-ray luminosities obtained by \citet{2020Ajello}, together with their fit to the data (orange line), that also includes the upper limits. For the SMC, LMC and M33, we show in addition to their FIR luminosity (in grey) also their SFRs as black points (see text for details).
Middle panels: contributions of neutral pion decay ($L_{\pi^0}$), IC emission ($L_{\mathrm{IC}}$) and bremsstrahlung ($L_{\mathrm{brems.}}$) to the total $\gamma$-ray luminosity $L_{\gamma}$, integrated over 0.1-100\,GeV, for $\zeta_{\mathrm{SN}}=0.05$ (left) and $\zeta_{\mathrm{SN}}=0.10$ (right) in our CR diffusion model. Lower panels: the calorimetric fraction $\eta_{\mathrm{cal,p}}$ (see Eq.~\ref{eq:eta_cal_p}) as a function of SFR of our simulated galaxies. The calorimetric limit, where hadronic losses dominate ($\eta_{\mathrm{cal,p}}\rightarrow 1$), is not reached by starburst galaxies.}
\label{fig:FIR-gamma-ray,fraction_Lgamma}
\end{figure*}

\begin{figure}
\includegraphics[]{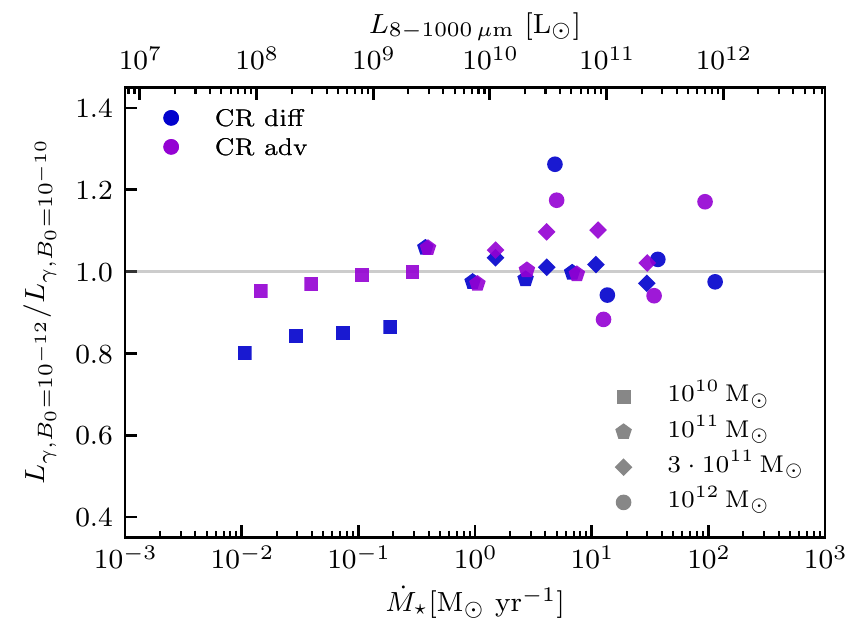}
\caption{In order to evaluate the effect of changing the initial magnetic field $B_0$ in our simulations with $\zeta_{\mathrm{SN}}=0.05$ on the resulting gamma-ray emission, we show the ratio of the gamma-ray luminosity from our runs adopting $B_0=10^{-12}\,\mathrm{G}$ relative to the gamma-ray luminosity from our runs with $B_0=10^{-10}\,\mathrm{G}$.} 
\label{fig:Lgamma_diff_B0}
\end{figure}

\subsubsection{Simulations}

If we sum up the integrated emission from 0.1 to 100 GeV in each cell of our simulated galaxies, we get the total $\gamma$-ray luminosity, with contributions from the different radiation processes discussed above. In the upper panels in Fig.~\ref{fig:FIR-gamma-ray,fraction_Lgamma} we show the resulting FIR-$\gamma$-ray relation for our simulated galaxies with different halo masses (corresponding to different symbols) at different times, which corresponds to different star formation rates in the simulations. Starting from the time of the peak of the SFR, we chose for each simulation the snapshots where the SFR has approximately decreased by an e-folding. We show the model that only accounts for CR advection (violet) in comparison to the model that additionally includes anisotropic CR diffusion with a constant diffusion coefficient $D=10^{28}\, \mathrm{cm^2\,s^{-1}}$ (dark blue). Additionally, we show simulations with a higher diffusion coefficient of $D=3\times 10^{28}\, \mathrm{cm^2\,s^{-1}}$ (light blue) for the smallest halo masses of $10^{10}\,\mathrm{M_{\odot}}$. On top of that, we plot the observations (black symbols) and the upper limits detected by \Fermi LAT from \citet{2020Ajello} and their best fit relation (orange line). We consider two different injection efficiencies $\zeta_{\mathrm{SN}}=0.05$ (top left-hand panel) and $\zeta_{\mathrm{SN}}=0.10$ (top right-hand panel). 

Furthermore, we note that the influence of the choice of initial magnetic field ($B_0=10^{-10}$ or $10^{-12}\,\mathrm{G}$) on the resulting $\gamma$-ray emission is only marginal.\footnote{Pictured in the upper panels of Fig.~\ref{fig:FIR-gamma-ray,fraction_Lgamma} are the simulations with $B_0=10^{-10}\,\mathrm{G}$ for $\zeta_{\mathrm{SN}}=0.05$ (left-hand panel) and $B_0=10^{-12}\,\mathrm{G}$ for $\zeta_{\mathrm{SN}}=0.1$ (right-hand panel).} This is because gravo-turbulence driven by the initial infall of gas in our simulations results in a turbulent, small-scale dynamo that exponentially amplifies the seed magnetic field so that it saturates at a level close to equipartition with the kinetic turbulence at small scales, from where it is further amplified and ordered on larger scales \citep{Pfrommer2021}. As a result, the time of magnetic saturation and the launching of galactic winds vary among those models with a different seed magnetic field. This is shown in Fig.~\ref{fig:Lgamma_diff_B0}, where we quantify this effect on the total gamma-ray luminosity from our simulations with $\zeta_{\mathrm{SN}}=0.05$. When changing the initial magnetic field from $B_0=10^{-10}$ to $10^{-12}\,\mathrm{G}$, the difference in gamma-ray luminosities is less than 27 per cent and is below 10 per cent for three quarters of all analysed snapshots.

Two conclusions can be drawn from the upper panels in Fig.~\ref{fig:FIR-gamma-ray,fraction_Lgamma}. First, we find that the smaller CR injection efficiency at SNR of $\zeta_{\mathrm{SN}}=0.05$ (upper left-hand panel) is preferred by the observations. Adapting a higher injection efficiency of $\zeta_{\mathrm{SN}}=0.10$ (upper right-hand panel) overestimates the observed relation for almost all combinations of halo masses and SFRs. Only the simulations with a higher diffusion coefficient of $D=3\times 10^{28}\, \mathrm{cm^2\,s^{-1}}$ manage to come close to the best-fit relation inferred from observations.
Furthermore, even for the lower injection efficiency, the simulations that only account for advection of CRs (purple symbols) increasingly deviate from the observed relation for decreasing SFRs, where they produce significantly larger $\gamma$-ray luminosities in comparison to the observed galaxies \citep[see also][]{2017bPfrommer}.

\begin{figure*}
\begin{centering}
\includegraphics[scale=1]{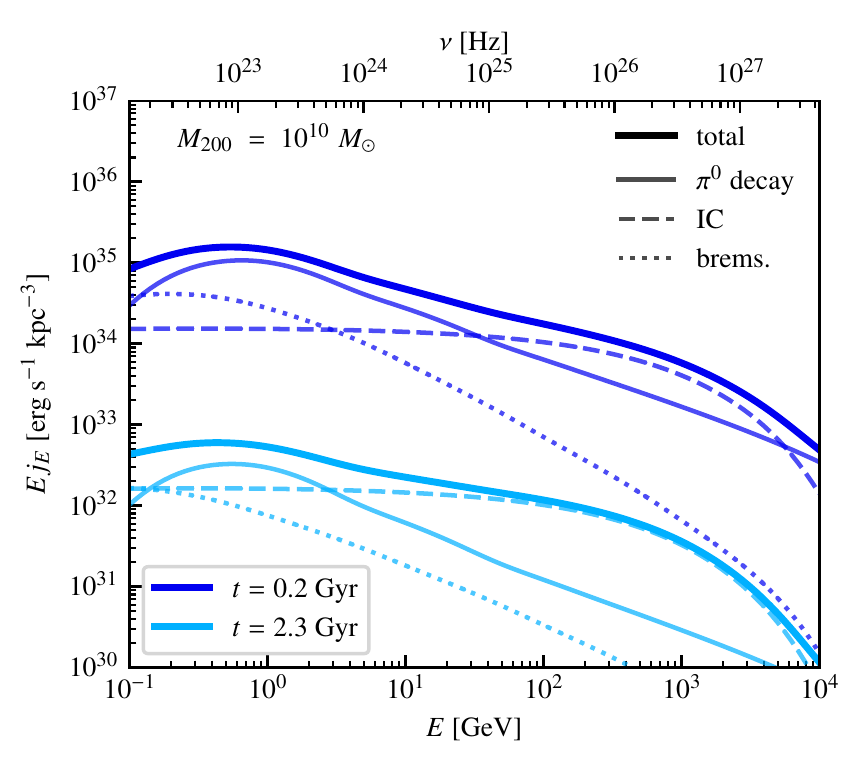}
\includegraphics[scale=1]{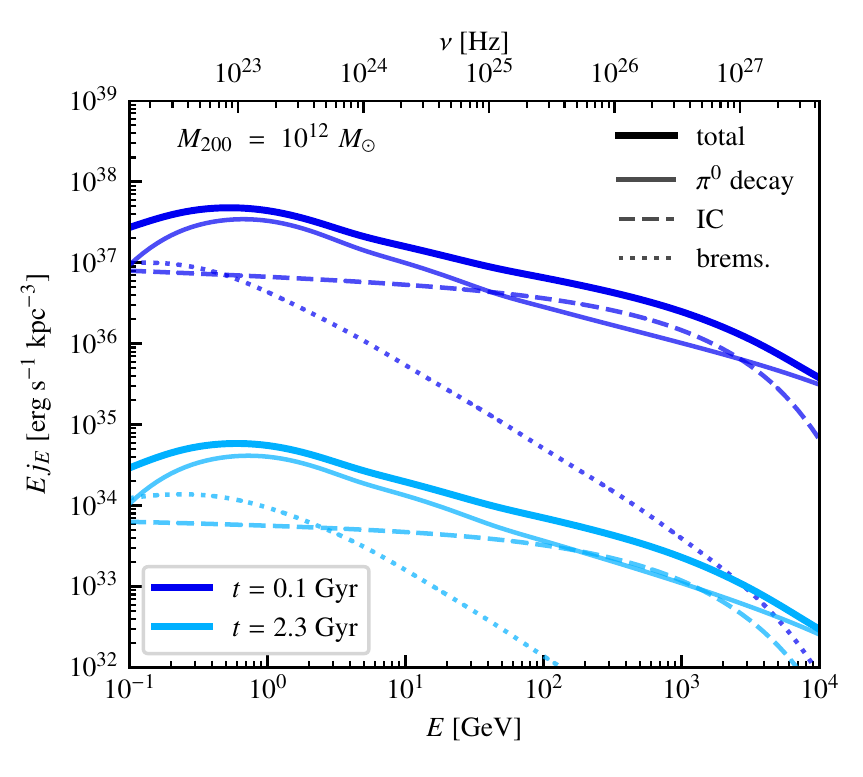}
\par\end{centering}
\caption{$\gamma$-ray spectra for halo masses $M_{200}=10^{10}$ (left-hand panel) and  $M_{200}=10^{12}$ (right-hand panel)  at the time of the peak of star formation and at a later time, when the SFR has decreased by two e-foldings. Shown are the contributions from neutral pion decay, IC and bremsstrahlung emission for each snapshot, as well as the total emitted spectrum (as indicated in the legend), respectively.} 
\label{fig:gamma-ray-spectra-1e12-1e10-times}
\end{figure*}

To explain these trends in our simulations, we dissect the different contributions to the total $\gamma$-ray luminosity in Fig.~\ref{fig:FIR-gamma-ray,fraction_Lgamma} (middle panels), where we show the fractional contributions of neutral pion decay, IC emission and bremsstrahlung emission to the total $\gamma$-ray luminosity $L_{\gamma}$ in our simulations for $\zeta_{\mathrm{SN}}=0.05$ and $0.10$. The contribution of neutral pion decay to the total $\gamma$-ray emission decreases towards lower SF galaxies and is filled in by IC emission from CR electrons, whose contribution reaches up to $\sim 40$ per cent in galaxies with a small SFR of $\sim 10^{-2}\,\mathrm{M_{\odot}\,yr^{-1}}$. The fractional contribution of bremsstrahlung contributes $\sim 20$ per cent to the gamma-ray luminosity across all SFRs.

The smaller contribution of hadronic emission at small SFRs can be attributed to the decreasing gas density in those smaller galaxies, leading to less efficient hadronic losses of CR protons and allowing escape losses to dominate \citep[see also e.g.\ ][]{2006Thompson, 2007Thompson, 2010Strong, 2010Lacki, 2011Lacki, 2014Martin, 2017bPfrommer}. This is underlined by Fig.~\ref{fig:FIR-gamma-ray,fraction_Lgamma} (lower panels), which we discuss in the following.


On average, a gamma-ray photon with an energy $E$ can be produced by a proton with a kinetic energy of $T_{\mathrm{p}}\approx 8E$. In order to quantify the energy fraction of CR protons that are able to produce gamma-rays in the energy band ranging from $E_1$ to $E_2$, we define the bolometric energy fraction
\begin{align}
\xi_\rmn{bol} = \frac{\varepsilon_{\mathrm{p}}(p_1,p_2)}{\varepsilon_{\mathrm{p}}(0,\infty)}&= \frac{\int_{p_1}^{p_2}q_\mathrm{p}(p_\mathrm{p})\,T_\mathrm{p}(p_\mathrm{p})dp_\mathrm{p}}{\int_0^{\infty}q_\mathrm{p}(p_\mathrm{p})\,T_\mathrm{p}(p_\mathrm{p})dp_\mathrm{p}} \approx 0.6,
\end{align}
where the normalised proton momenta $p_1$ and $p_2$ are given by $p_{1,2}=\sqrt{[8E_{1,2}/(m_\mathrm{p}c^2) +1]^2 -1}$ and we adopt $E_1=0.1$~GeV and $E_2=100$~GeV.
This allows us to define the calorimetric fraction as the ratio of the luminosity of all pion-decay end products produced in hadronic collisions ($L_\pi$) to the proton luminosity ($L_{\mathrm{p}}$) as 
\begin{align}
\eta_{\mathrm{cal,p}}
= \frac{L_{\pi}}{\xi_\rmn{bol}L_{\mathrm{p}}}
\approx 1.7\times \frac{\sum_i L_{\pi, i}}{\sum_i L_{\mathrm{p},i}}.
\label{eq:eta_cal_p}
\end{align}
We estimate the total pion luminosity in each cell from the gamma-ray luminosity resulting from neutral pion decay, i.e.\ $L_{\pi,i} \approx 3L_{\pi^0,i}=3L_{\gamma,i}$.
The injected proton luminosity $L_{\mathrm{p},i}$ is computed from the SFR in each cell $i$ via
\begin{align}
L_{\mathrm{p},i}=\zeta_{\mathrm{SN}} \dot{M}_{\star,i} \,\epsilon_{\mathrm{SN}} \ ,
\end{align}
where $\epsilon_{\mathrm{SN}}=E_{\mathrm{SN}}/M_{\star} = 10^{51}\,\mathrm{erg}/ (100\mathrm{M_\odot})= 10^{49}\,\mathrm{erg\,M_{\odot}^{-1}}$ quantifies the SN energy release per unit mass, under the assumption of a \cite{2003Chabrier} initial mass function and assuming that stars with a mass above 8~$\mathrm{M_\odot}$ explode as SNe.


Hence, in the calorimetric limit, where hadronic losses dominate over escape losses, we obtain $\eta_{\mathrm{cal,\,p}}\to 1$. We find that highly SF galaxies with $\mathrm{SFR} \gtrsim 1\,\mathrm{M_{\odot}\,yr^{-1}}$ approach this limit, but level off at around 0.7. On average, we find that SF galaxies with SFRs ranging from $\sim$1 to 100~$\mathrm{M_\odot\,yr^{-1}}$ exhibit calorimetric fractions between 0.3 to 0.7. For starburst galaxies like NGC~253 and M82, this is roughly consistent with what has been previously found by \cite{2011Lacki}, who estimated calorimetric fractions of NGC~253 and M82 ranging from about 0.2 to 0.4.
Even though this means that highly SF galaxies lose a significant amount of energy due to hadronic interactions, we note that a calorimetric fraction of e.g., $\sim$40 per cent implies that the remaining 60 per cent of CR energy diffuses out of SF regions and is available for CR feedback in form of CR-driven galactic winds. In normally SF galaxies with $\mathrm{SFR} \lesssim 1\,\mathrm{M_{\odot}\,yr^{-1}}$, losses due to CR diffusion start to become dominant over hadronic losses, which leads to a decreasing calorimetric fraction. Consequently, the total $\gamma$-ray emission of these galaxies falls short of the calorimetric relation shown in the upper panels of Fig.~\ref{fig:FIR-gamma-ray,fraction_Lgamma} (dashed green line).  Modeling a larger CR diffusion coefficient (of $D=3\times 10^{28}\, \mathrm{cm^2\,s^{-1}}$, light blue symbols) thus results in an even larger deviation from calorimetry.

\subsection{$\gamma$-ray spectra}  \label{subsec: gamma-ray spectra}

In addition to the spatial information and the total $\gamma$-ray luminosity of our steady-state models of CR electrons and protons, we study the $\gamma$-ray spectra of our simulated galaxies and compare them to observational data.

\subsubsection{Simulated $\gamma$-ray spectra}

\begin{figure*}
\begin{centering}
\includegraphics[scale=1.1]{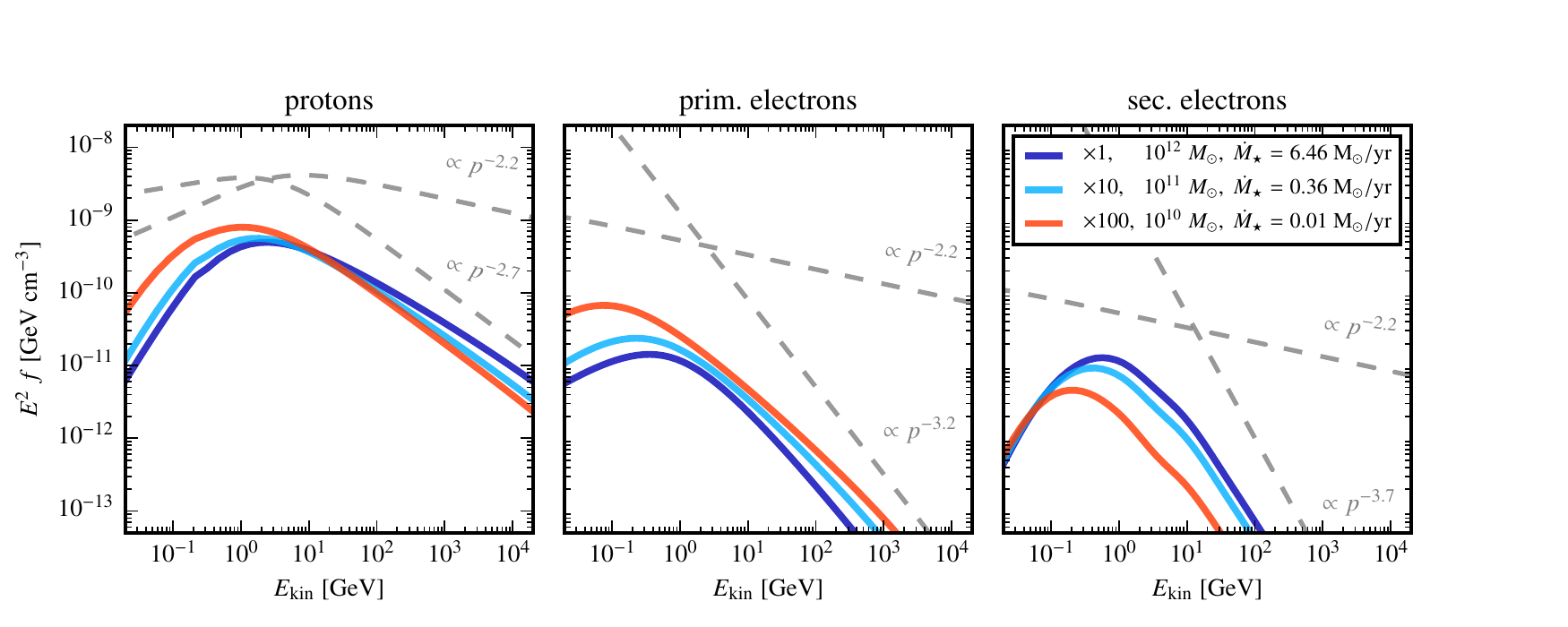}
\includegraphics[scale=1.1]{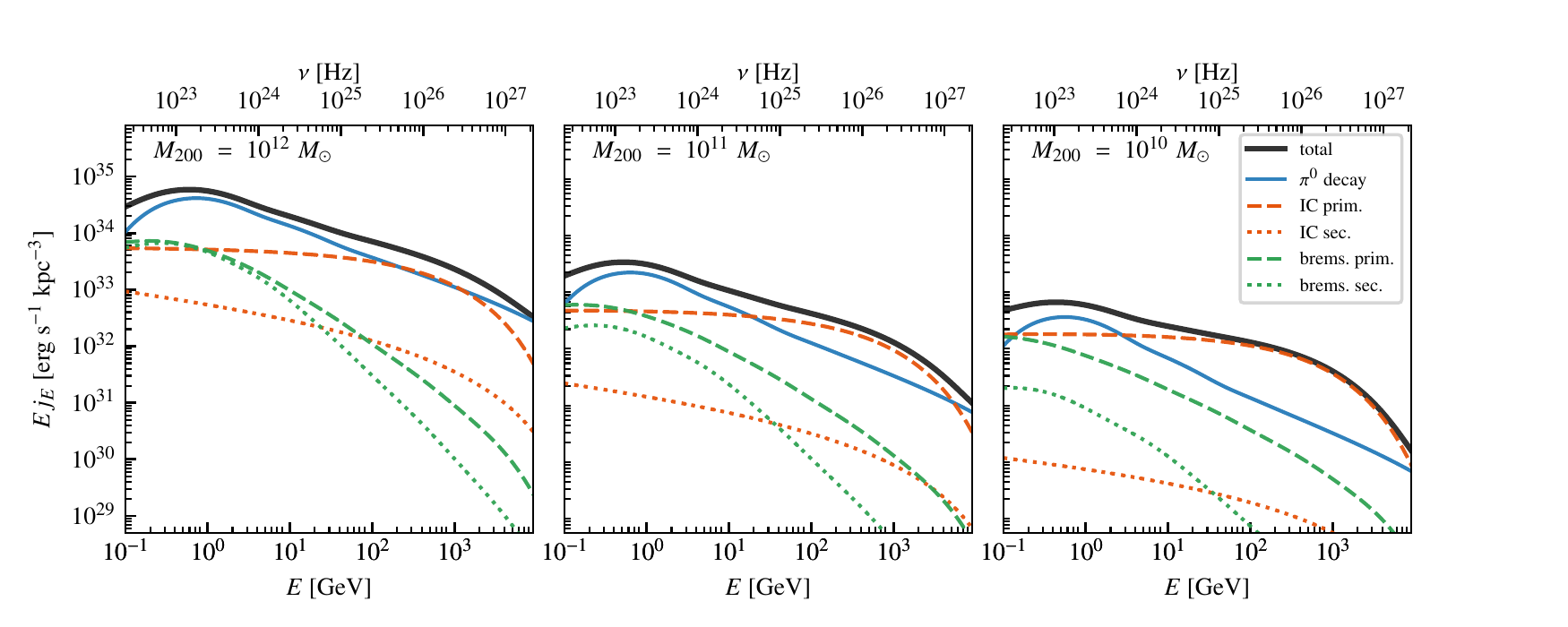}
\par\end{centering}
\caption{Upper panel: CR proton, primary and secondary electron spectra for three different halo masses, $10^{12}$, $10^{11}$ and $10^{10}\,\mathrm{M_\odot}$, all at $t=2.3\, \mathrm{Gyr}$, averaged over the gas scale-height and the radius that includes 99 per cent of the gamma-ray emission. For visual purposes, the CR spectra of the lower-mass halos are re-scaled as indicated in the legend.
Lower panels: resulting $\gamma$-ray spectra of the same halos as shown above. The contribution to the IC and bremsstrahlung arising from the primary (dashed lines) and secondary (dotted lines) electron populations are shown separately.} 
\label{fig:gamma-ray-spectra-3halos}
\end{figure*}

\begin{figure*}
\begin{centering}
\includegraphics[scale=1]{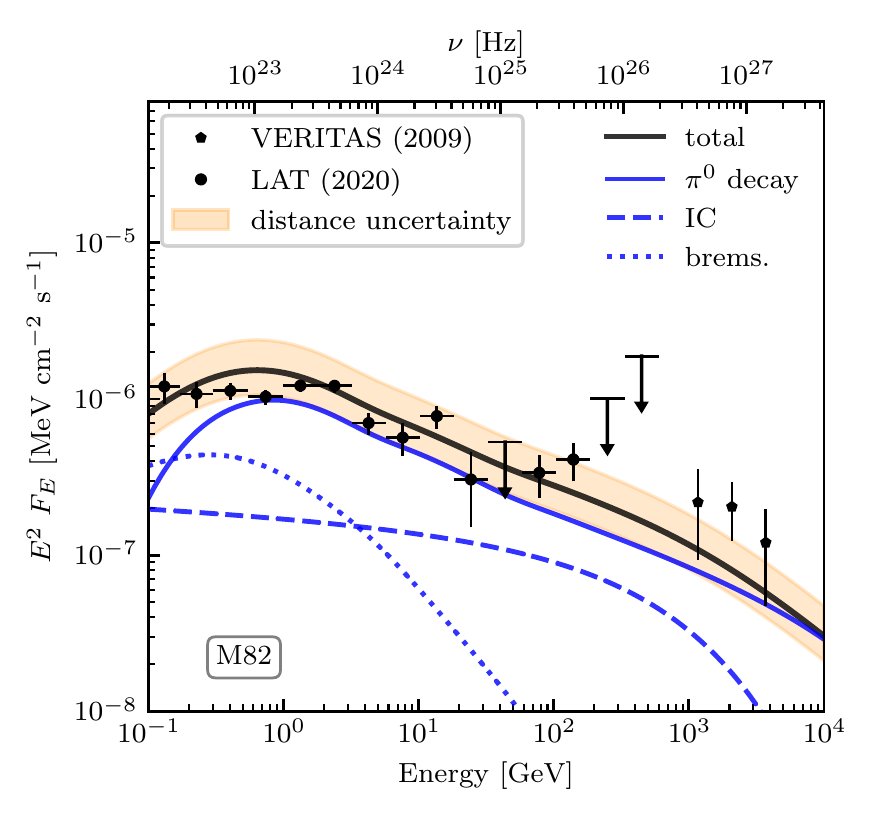}\includegraphics[scale=1]{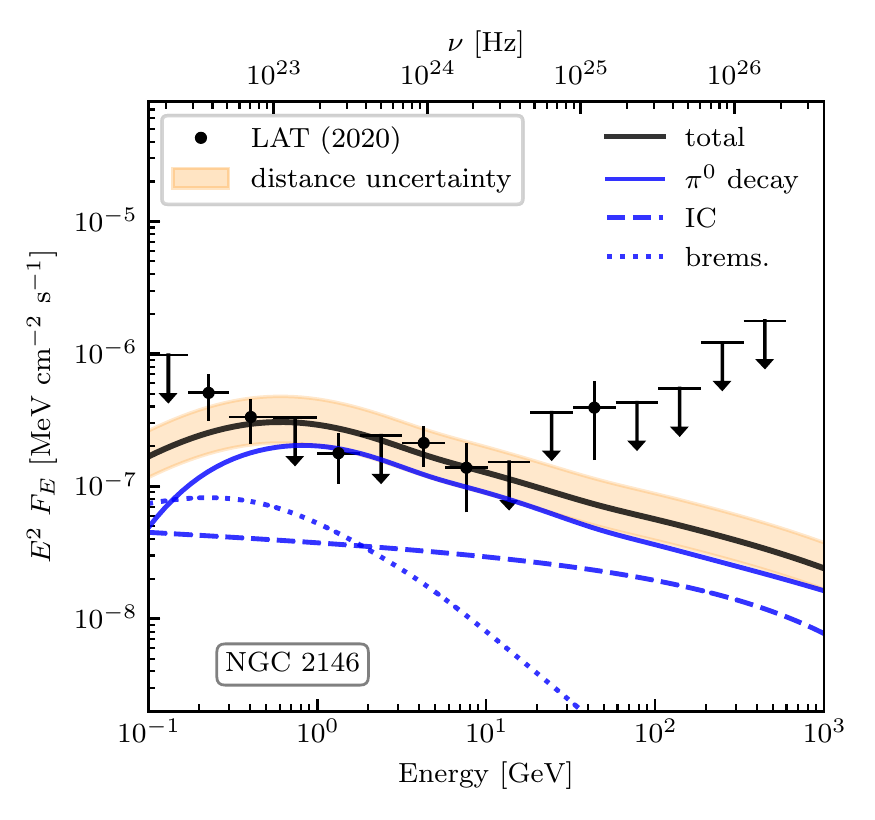}
\par\end{centering}
\begin{centering}
\includegraphics[scale=1]{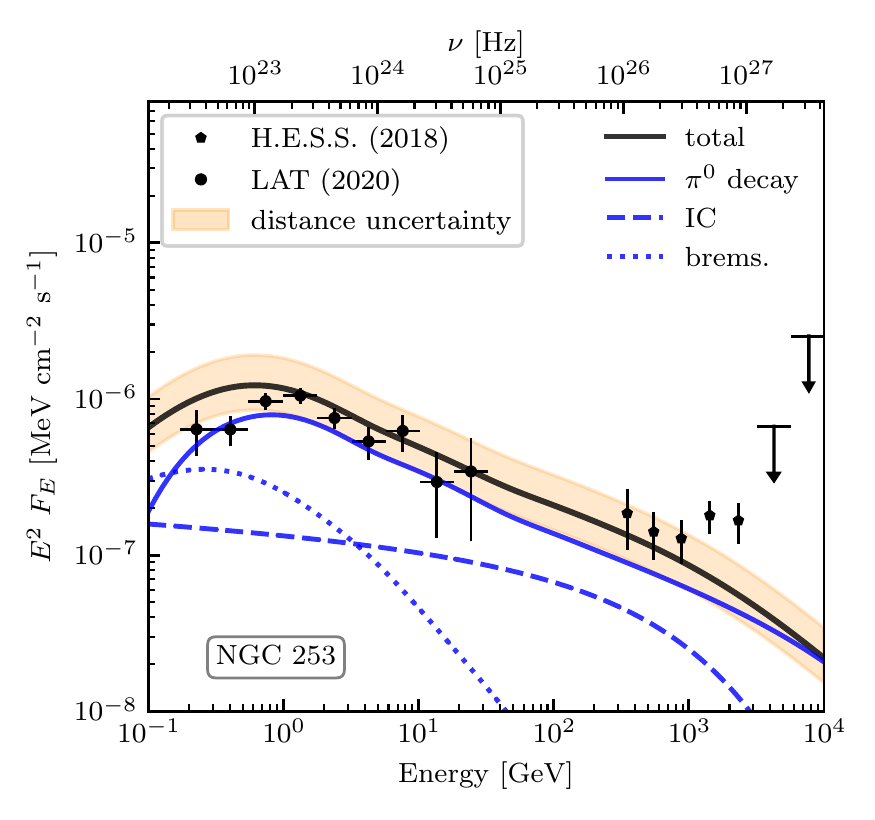}\includegraphics[scale=1]{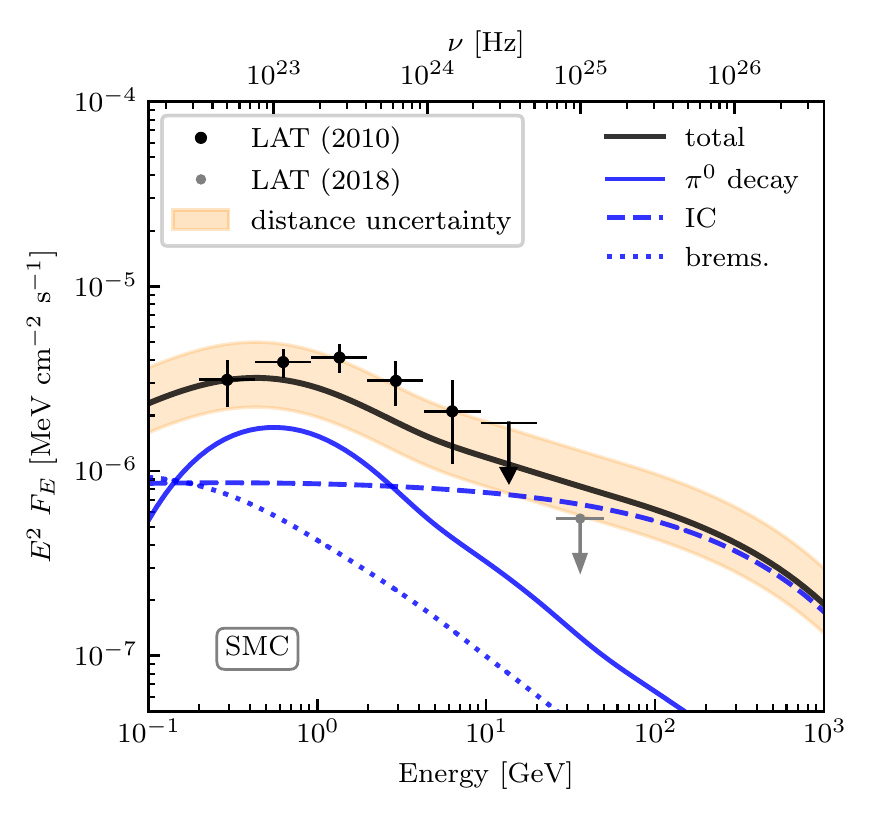}
\par\end{centering}
\caption{$\gamma$-ray spectra of the observed galaxies M82, NGC~2146, NGC~253 and SMC together with the emission of our simulated galaxies, which exhibit a similar total $\gamma$-ray luminosity and SFR. The simulated spectra are re-normalised to the corresponding observed total $\gamma$-ray luminosities for visual purposes (i.e.\ by factors of 0.71 to 1.03; see Table~\ref{Table-Galaxies}).}
\label{fig:gamma-ray-spectra-data}
\end{figure*}
\begin{figure}
\begin{centering}
\includegraphics[scale=1]{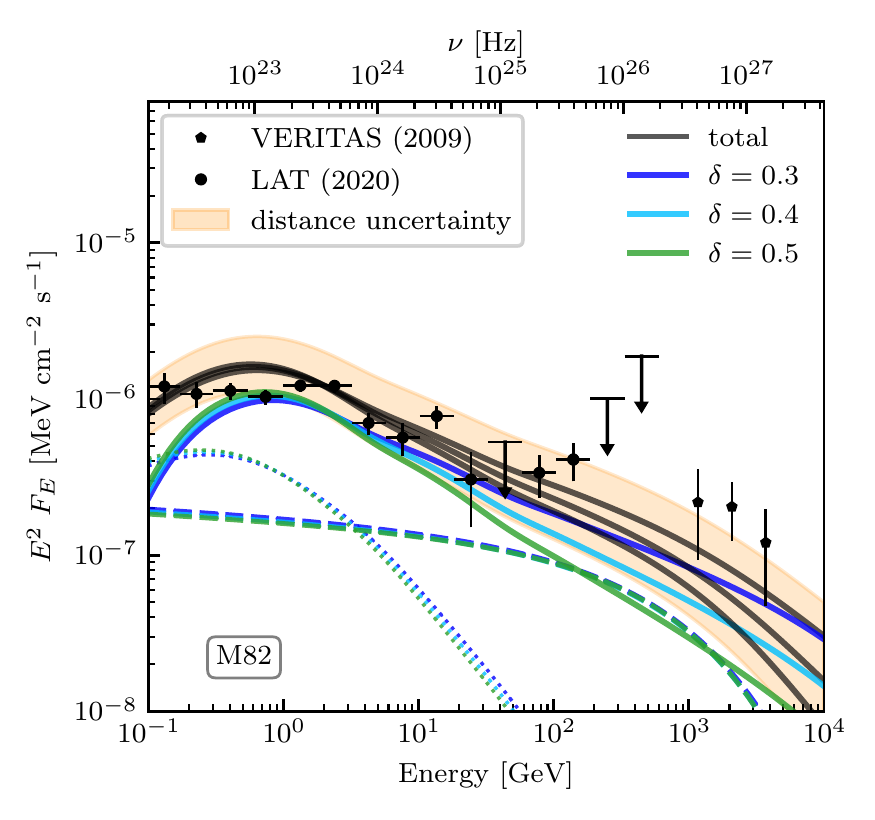}
\par\end{centering}
\caption{$\gamma$-ray spectrum of M82 (same as Fig.~\ref{fig:gamma-ray-spectra-data}, upper left panel), together with our simulated spectra for three different values of the energy dependence of diffusion timescale, i.e. $t_{\mathrm{diff}} \propto E^{-\delta}$. A higher value of $\delta$ implies faster diffusion at higher energies, which results in steeper spectra of CR proton and $\gamma$ rays from neutral pion decay.} 
\label{fig:gamma-ray-spectra-deltas}
\end{figure}

First, we investigate the effect of halo mass and temporal evolution of our simulations on the $\gamma$-ray spectra. In Fig.~\ref{fig:gamma-ray-spectra-1e12-1e10-times}, we show the $\gamma$-ray spectra of two different halo masses $M_{200}=10^{10}$ and $10^{12}\,\mathrm{M_{\odot}}$ at two different times, respectively. The time of the first snapshot is chosen at the time of the peak of the SF history, i.e., after 0.2 and 0.1\,Gyr, respectively. At the second time shown here, the SFR has decreased by two e-foldings, i.e., at 2.3\,Gyr in both cases. As the SFR decreases with time, the CR injection and thus its energy density drops, too. As a result, the total emitted spectrum is shifted downwards with time for both halo masses. While the IC emission in the smaller halo dominates the $\gamma$-ray emission above 40\,GeV at early times and above a few GeV at later times, its contribution in the more massive halo decreases over time. In all cases, the bremsstrahlung emission plays a subdominant role above energies of 1\,GeV and only becomes relevant at lower energies.

In Fig.~\ref{fig:gamma-ray-spectra-3halos} we assess the interplay between our steady-state CR spectra and the resulting $\gamma$-ray emission spectra. In the upper panels, we show the spectra of CR protons, primary and secondary electrons for three different halo masses ($M_{200}=10^{12},\,10^{11}$ and $10^{10}\,\mathrm{M_{\odot}}$) at the same time ($t=2.3\,\mathrm{Gyr}$), which corresponds to SFRs of 6.48, 0.36 and 0.01\,$\mathrm{M_{\odot}~yr}^{-1}$, respectively. The CR spectra are averaged over the gas scale-height, where the gas density has dropped by an e-folding (ranging from 0.6 to 0.9 kpc), and the radius, where 99 per cent of the gamma-ray luminosity is included. These range from $r=9.5$~kpc in the low-mass halo to $r=20.5$~kpc and $r=27.5$~kpc in the middle and high-mass halos, respectively. Furthermore, the CR spectra of the lower mass halos are re-scaled by factors of 10 and 100 (as indicated in the legend) for visual purposes, which enables us to identify the differences in their spectral slopes. In the case of CR protons, smaller galaxies with lower SFRs exhibit steeper CR proton spectra, indicating that diffusive losses (that are assumed to be energy dependent) become increasingly important, which is consistent with our findings of Fig.~\ref{fig:FIR-gamma-ray,fraction_Lgamma} (bottom panels). Consequently, their CR proton spectra steepen by 0.5 due to energy dependent diffusion, which yields a spectral index of 2.7. This also results in a steeper $\gamma$-ray spectrum resulting from neutral pion decay with decreasing SFR (lower panels of Fig.~\ref{fig:gamma-ray-spectra-3halos}).

By contrast, the primary CR electron spectrum steepens with higher SFR. This can be attributed to higher radiative losses of CR electrons under these conditions, which leads to a spectral index of the CR electron spectrum that is steeper by unity in comparison to the injected index, so that the cooled spectral index approaches 3.2. 
As radiation dominates over the magnetic energy density in most of the galaxy except for the very central regions of the galaxy (see Fig.~\ref{fig:Maps-properties}), CR electrons mainly cool via IC interactions.

The contribution of secondary electrons decreases with lower SFR as a consequence of the associated lower gas density. This is visible in the normalization of the spectra of secondary CR electrons in comparison to that of primary electrons, as well as the secondary bremsstrahlung and IC emission in comparison to the primary contributions, respectively. Hence, in the $10^{12}\,\mathrm{M_{\odot}}$ halo, which exhibits a higher SFR, secondary and primary bremsstrahlung contribute nearly equally. On the contrary, secondary bremsstrahlung is negligible in the $10^{10}\,\mathrm{M_{\odot}}$ halo. Similarly, the relevance of secondary IC emission decreases significantly from the $10^{12}$ to the $10^{10}\,\mathrm{M_{\odot}}$ halo.
Furthermore, the secondary electron spectra are steeper in comparison to the primary ones because they result from the steady-state CR proton population, that exhibit spectral indices of $2.2<\alpha_{\mathrm{p}}<2.7 $ due to energy dependent diffusion losses. After having cooled, secondary electron spectra are thus steepening further and approach spectral indices of $\alpha_{\mathrm{e,sec}} \gtrsim 3.2$, whereas the primary electron spectrum steepens at most by unity so that $\alpha_{\mathrm{e,prim}} \lesssim 3.2$. Overall, this leads also to steeper IC and bremsstrahlung spectra from the secondary CR electron population in comparison to primary electrons.

\subsubsection{Comparison to observed $\gamma$-ray spectra}

Looking at the upper panels in Fig.~\ref{fig:FIR-gamma-ray,fraction_Lgamma}, we can easily find suitable snapshots of our simulations that correspond to individual observed galaxies in terms of their total $\gamma$-ray luminosity and their SFR. We consider the four galaxies NGC~2146, NGC~253, M82 and the SMC. In Table~\ref{Table-Galaxies}, we identify the corresponding snapshots that resemble the observed galaxies in terms of their SFR and total $\gamma$-ray luminosity, using our simulations  with $\zeta_{\mathrm{SN}}=0.05$ and $B_0=10^{-10}\,\mathrm{G}$. Even though the initial magnetic field does not influence the resulting $\gamma$-ray emission, we chose a value of $B_0=10^{-10}\,\mathrm{G}$ in order to better reproduce the FIR-radio relation, which we discuss in \citetalias{2021WerhahnIII}. The $\gamma$-ray spectra of the individual galaxies are shown in Fig.~\ref{fig:gamma-ray-spectra-data}, together with observations by LAT \citep{2020Ajello}. For M82, we additionally show \citet{2009VERITAS_M82} data, for NGC~253 \citet{2018HESS_NGC253} data, and for the SMC, we show \Fermi-LAT data \citep{2010Abdo, Lopez_2018}. The distances to the galaxies are taken from \citet{2020Ajello} and references therein. The distance uncertainties are indicated by the orange band and are assumed to be 20 per cent, because typical distance uncertainties for galaxies within $\sim$ 25~Mpc are around 10-20 per cent \citep{2001Freedman}.

In case of starburst galaxies, i.e.\ for NGC~2146, NGC~253 and M82, we successfully match the observations at low energies. Only in the case of NGC~253, we slightly over-predict the data at 200-300 MeV which may indicate a too large leptonic bremsstrahlung contribution. In order to better match the data in the very high energy regime in these starburst galaxies, we need to assume the energy dependence of the diffusion timescale to scale as $E^{-\delta}$ with $\delta=0.3$. This corresponds to a turbulent Kolmogorov spectrum \citep{2007Strong}. The effect of varying $\delta$ on the $\gamma$-ray spectrum is depicted in Fig.~\ref{fig:gamma-ray-spectra-deltas}, where we show the $\gamma$-ray spectra for our M82 analogue adapting $\delta=\{0.3,0.4,0.5\}$. As expected, a higher value of $\delta$ leads to a steeper spectrum of CR protons and of the resulting neutral pion decay emission.
In case of the SMC, we stick to our model with $\delta=0.5$, which leads to a steeper spectrum at high energies that better agrees with the upper limit found by \citet{2010Abdo}. However, we do not match the upper limit determined by \citet{Lopez_2018}. 
The resulting $\gamma$-ray luminosities given in Table~\ref{Table-Galaxies} are calculated using $\delta = 0.3$ for NGC~2146, NGC~253 and M82 and are only $3-6$ per cent smaller, if we adopt $\delta = 0.5$.

\section{Discussion and Conclusions}  \label{sec: Discussion and conclusion}

In this paper, we evaluate the non-thermal $\gamma$-ray emission from MHD simulations of isolated galaxies with different halo masses. The emission processes include neutral pion decay resulting from hadronic interactions of CR protons with the ISM, as well as IC and bremsstrahlung emission from CR electrons. We model the spectra of CR protons, primary and secondary electrons with a cell-based steady-state approximation in post-processing of our simulations, following the approach detailed in \citetalias{2021WerhahnI}. This enables us to produce $\gamma$-ray maps of the emission processes as well as their energy spectra. Furthermore, we calculate a total $\gamma$-ray luminosity for each simulated galaxy and find that we better reproduce the observed correlation with the FIR luminosity if we adopt a low injection efficiency of CRs at SNe of $\zeta_{\mathrm{SN}}=0.05$. Furthermore, the simulations only accounting for advection of CRs clearly overproduce the observed $\gamma$-ray luminosities at small SFRs and halo masses. Only the models with anisotropic CR diffusion agree with the observed deviation from calorimetry with decreasing SFRs.

Despite current belief, we find that the leptonic contribution to the total $\gamma$-ray emission in form of bremsstrahlung and IC emission is not negligible. While it is subdominant in comparison to the pion decay $\gamma$ rays in starburst galaxies, it becomes progressively more important towards more moderate SF (and smaller) galaxies. On the other hand, this is also the regime where our assumption underlying the IC emission is weak, i.e.\ that the interstellar radiation field is dominated by the reprocessed UV light of young stellar populations, that gets re-emitted in the FIR by dust. Still, independent of the exact modeling of the incident radiation field for the IC emission, the pionic $\gamma$-ray emission in our model is not able to account for both the observed $\gamma$-ray luminosities as well as the spectral shapes of galaxies with low SFRs such as the SMC alone. As diffusive losses become more relevant in these low-density galaxies, this leads to a steepening of the CR proton spectra (depending on the exact energy dependence incorporated in the diffusion timescale). This necessarily results in steeper pion decay $\gamma$-ray spectra, which would be in conflict with the observations if this were the only relevant emission channel. Hence, a leptonic contribution to the $\gamma$-ray emission from IC scattering is indispensable in our models.

One of the arguments made in the literature in order to conclude a sub-dominant leptonic contribution to the total $\gamma$-ray emission depends on several underlying assumptions, that are only valid in the nuclear regions of starbursts, i.e.\ for high gas densities, magnetic field strengths and wind velocities.
The magnetic field strength is either deduced from an energy equipartition argument, or from fitting several free parameters in one-zone models \citep[see e.g.,][]{2004Torres,2010Lacki,2011Lacki,2013Yoast-Hull} or axisymmetric two-dimensional models \citep{2020Buckman} in order to simultaneously match the observations. The former approach assumes that the kinetic energy density $\varepsilon_{\mathrm{kin}} = \rho\,v^2/2$ is equal to the magnetic energy density, $\varepsilon_{\mathrm{mag}} = B^2/(8\upi)$. Assuming turbulent velocities of $\sim 20\,\mathrm{km/s}$ and gas densities of $n\sim 100\,\mathrm{cm^{-3}}$ yields in this approach magnetic field strengths of the order of $100\,\mathrm{\umu G}$. The fast cooling of CR electrons due to synchrotron emission in these strong magnetic fields has been proposed to be one of the reasons for suppressing the possible leptonic contribution to the $\gamma$-ray emission \citep{2011Lacki}. In contrast, $\gamma$-ray observations of the Galactic center suggest that CRs might not be able to penetrate into the densest regions and thus may be far from calorimetry \citep{2011bCrocker,2011aCrocker}.

However, those extreme properties are at most reached in the very central region of our simulated galaxies. Additionally, the energy density of the interstellar radiation field (or its equivalent magnetic field strength, see Fig.~\ref{fig:Maps-properties}, upper panels) persists throughout the whole galaxy. The cooling of the electrons is hence not fully dominated by synchrotron cooling, enabling IC losses to be non-negligible, except in the very central regions of starburst galaxies. In particular, in the low density interstellar medium of dwarfs, we find that the IC emission might be able to contribute up to 40 per cent of the total $\gamma$-ray luminosity, as can be seen in Fig.~\ref{fig:FIR-gamma-ray,fraction_Lgamma} (middle panels).
Still, our models of starburst galaxies M82 and NGC~253 (that are chosen to match the observed SFRs and $L_{\gamma}$) also match their observed radio luminosities at 1.4 GHz. We will address the radio emission of our simulated galaxies in \citetalias{2021WerhahnIII} in more detail.

Our findings suggest that SF galaxies are not reaching the calorimetric limit, and depart from this limit even further with smaller SFR and halo masses, which is due to the increasing relevance of diffusion losses. Hence, the contribution of neutral pion decay to the total $\gamma$-ray luminosity decreases at low SFRs, which is partly compensated by the larger contribution of leptonic IC $\gamma$-ray emission. Furthermore, the departure from calorimetry implies that there is still a considerable amount of CR energy left for feedback processes, even in highly SF galaxies.

\section*{Acknowledgments}

We thank our referee for a careful reading and an insightful report that helped to improve the paper. MW, CP, and PG acknowledge support by the European Research Council under ERC-CoG grant CRAGSMAN-646955. This research was supported in part by the National Science Foundation under Grant No.\ NSF PHY-1748958.

\section*{Data Availability}
 
The simulations and data analysis scripts underlying this article will be shared on reasonable request to the corresponding author. The \arepo code is publicly available.

\bibliographystyle{mnras}
\bibliography{literatur}

\appendix

\section{Radiative processes in the $\gamma$-ray regime} \label{sec: Radiation processes}
We describe in the following the relevant processes in the $\gamma$-ray regime. In this context, we use the following definitions.
The production spectrum of $N_\gamma$ photons or source function $q_{\gamma}$ is defined in units of $\mathrm{ph~erg^{-1}\,s^{-1}\,cm^{-3}}$ as
\begin{align}
q_{\gamma}=\frac{\mathrm{d}N_{\gamma}}{\mathrm{d}E\mathrm{d}t\mathrm{d}V},
\end{align}
where $E$ is the energy of the emitted photon, $t$ denotes the unit time and $V$ is the unit volume.
The production spectrum is connected to the different definitions of the emissivities via
\begin{align}
j_{E}&=E\frac{\mathrm{d}N_{\gamma}}{\mathrm{d}E \mathrm{d}V \mathrm{d}t}\\
j_{\nu}&=E\frac{\mathrm{d}N_{\gamma}}{\mathrm{d}\nu \mathrm{d}V \mathrm{d}t}=hj_{E},
\label{eq:j_E, j_nu}
\end{align}
where $h$ denotes Planck's constant. The total luminosity in erg~s$^{-1}$ is obtained by integrating the emissivity from energy $E_{1}$ to $E_{2}$ and over the total source volume $\Omega$, i.e.,
\begin{align}
L_{E_{1}-E_{2}}=\intop_\Omega \mathrm{d}V \intop_{E_{1}}^{E_{2}} \mathrm{d}E\,E\,q_{\gamma}.
\end{align}
Observing the emitting object from a luminosity distance $d$ yields an observed spectral flux
\begin{align}
F_{E}=\frac{1}{4\upi d^2}  \intop_\Omega \mathrm{d}V\,q_{\gamma}
\end{align}
in units of $\mathrm{ph~erg^{-1}\,s^{-1}\,cm^{-2}}$.

\subsection{$\gamma$-ray emission from neutral pion decay\label{subsec:Gamma-ray-emission-from neutral pion decay}}

The collisions of CR protons with protons and other nuclei in the ambient interstellar medium give rise to the production of several secondary particles. In particular, inelastic proton-proton (pp) collisions produce mainly pions, that lead to the production of $\gamma$ rays, secondary electrons/positrons and neutrinos:
\begin{eqnarray}
  \pi^\pm &\rightarrow& \mu^\pm + \nu_{\mu}/\bar{\nu}_{\mu} \rightarrow
  e^\pm + \nu_{e}/\bar{\nu}_{e} + \nu_{\mu} + \bar{\nu}_{\mu},\nonumber\\
  \pi^0 &\rightarrow& 2 \gamma \,.\nonumber
\end{eqnarray}
These secondary particles contribute to the leptonic radiation processes in addition to the primary electrons, that will be discussed in Sections~\ref{subsec:IC-emission} and \ref{subsec:Bremsstrahlung}.

The source function $q_{\gamma}$ that results from the decay of neutral pions following a pp collision is given by
\begin{equation}
q_{\gamma}(E) = c n_{\mathrm{H}} \intop_{E_{\mathrm{p},\mathrm{min}}}^{\infty} \mathrm{d}E_{\mathrm{p}} 
f_{\mathrm{p}} (E_{\mathrm{p}} )\frac{\mathrm{d}\sigma_{\gamma}(E,E_{\mathrm{p}})}{\mathrm{d}E},
\label{eq:production of gamma-rays}
\end{equation}
where $f_{\mathrm{p}}$ denotes the CR proton distribution, $E_{\mathrm{p}}$ is the total proton energy, $m_{\mathrm{p}}$ is the proton mass and $c$ the speed of light. Furthermore, $E$ is the energy of the emitted $\gamma$-ray photon and $n_{\mathrm{H}}$ the hydrogen number density. It yields an emissivity that we denote by $j_{\nu,\pi^0}=Eq_{\gamma}$. 
In the following, we denote neutral pions with $\pi$, if not stated otherwise.
The differential cross section of $\gamma$-ray production is given by 
\begin{equation}
\frac{\mathrm{d}\sigma_{\gamma}(E,E_{\mathrm{p}})}{\mathrm{d}E} = 2 \intop_{E_{\pi,\mathrm{min}}}^{E_{\pi,\mathrm{max}}}\mathrm{d}E_{\pi}\frac{\mathrm{d}\sigma_{\pi}(E_{\mathrm{p}},E_{\pi})}{\mathrm{d}E_{\pi}}f_{\gamma,\pi}(E,E_{\pi}).
\label{eq:dsigma_gamma/dE_gamma}
\end{equation}
Here, the normalised energy distribution $f_{\gamma,\pi}(E,E_{\pi})$ gives the probability of the production of a $\gamma$-ray photon with energy $E$ from a single pion energy $E_{\pi}$ and the factor of 2 accounts for the decay of one neutral pion into two $\gamma$ rays. Following e.g. \citet{1971NASSP.249.....S}, the Green's function for neutral pion decay is 
\begin{equation}
f_{\gamma,\pi}(E,E_{\pi}) =\frac{1}{\sqrt{E_{\pi}^{2}-m_{\pi}^{2}c^{4}}},\label{eq: pion decay greens function}
\end{equation}
where $m_\pi$ denotes the rest mass of neutral pions. From investigating the relativistic kinematics of the pp-collision, one can find the proton's threshold of pion production \citep[see e.g. ][]{1994A&A...286..983M}. If the kinetic energy of the proton in the center-of-momentum system is larger than the pions rest mass energy, a pion can be created. Transforming this requirement back to the lab system yields the threshold energy for pion production. It reads
\begin{align}
\frac{E_{\mathrm{p},\mathrm{min}}}{m_{\mathrm{p}}c^{2}}
=2\left[1+\frac{m_{\pi}}{2m_{\mathrm{p}}}\right]^{2}-1 = 1.22\,\frac{\mathrm{GeV}}{m_{\mathrm{p}}c^{2}}
\end{align}
and corresponds to the lower limit of the integral in Eq.~\eqref{eq:production of gamma-rays}. Following \citet{2014PhRvD..90l3014K}, in order to correctly define the limits of the integral in Eq.~(\ref{eq:dsigma_gamma/dE_gamma}), we consider the following quantities.
First, the total energy and momentum of the pion in the center-of-mass (CM) system are given by
\begin{align}
E_{\pi,\mathrm{CM}}=\frac{s-4m_\mathrm{p}^2c^4-m_\pi^2c^4}{2 \sqrt{s}}
\end{align}
and
\begin{align}
P_{\pi,\mathrm{CM}} = \sqrt{(E_{\pi,\mathrm{CM}})^2 -m_\pi^2 c^4}/c
\end{align}
where $s=2m_\mathrm{p}c^2(T_\mathrm{p}+2m_\mathrm{p}c^2)$ is the squared center-of-mass energy. From this, the Lorentz factor and velocities of the CM system are given by
\begin{align}
\gamma_\mathrm{CM}=\frac{T_\mathrm{p}+2m_\mathrm{p}c^2}{\sqrt{s}}
\end{align}
and $\beta_\mathrm{CM}=\sqrt{1-\gamma_\mathrm{CM}^{-2}}$.
The maximum allowed energy for the created pion given in the lab frame is
\begin{align}
E_{\pi,\mathrm{max}}=\gamma_{\mathrm{CM}}(E_{\pi,\mathrm{CM}}+cP_{\pi,\mathrm{CM}} \beta_{\mathrm{CM}})
\end{align}
and corresponds to the upper limit in Eq.~(\ref{eq:dsigma_gamma/dE_gamma}).
The corresponding Lorentz factor and velocity are denoted by
$\gamma_{\pi,\mathrm{LAB}}=E_{\pi,\mathrm{max}}/(m_\pi c^2)$ and $\beta_{\pi,\mathrm{CM}}=\sqrt{1-(\gamma_{\pi,\mathrm{CM}})^{-2}}$.

To ensure that the photon energy is $E<E_{\max}$, i.e. the maximum allowed energy for the photon from kinematic considerations, we define 
\begin{align}
Y_\gamma = E+\frac{m_{\pi}^{2}c^{4}}{4E},\\
Y_{\gamma,\mathrm{max}} = E_{\mathrm{max}}+\frac{m_{\pi}^{2}c^{4}}{4E_{\mathrm{max}}}
\end{align}
and 
\begin{align}
X_\gamma = \frac{Y_\gamma - m_\pi c^2}{Y_{\gamma,\mathrm{max}} - m_\pi c^2}.
\end{align}
Then, we require $0<X_\gamma<1$.
Additionally, the lower limit of the integral in Eq.~(\ref{eq:dsigma_gamma/dE_gamma}), $E_{\pi,\mathrm{min}}(E)$ is the minimum energy that is needed to produce a photon of energy $E$, i.e.
\begin{equation}
E_{\pi,\mathrm{min}}=\max \left(m_{\pi}c^2, E+\frac{m_{\pi}^{2}c^{4}}{4E}\right).\label{eq:E_pi_min (E_gamma)}
\end{equation}

There are different parametrizations for the required terms in Eq.~\eqref{eq:production of gamma-rays} in the literature, valid in different energy ranges. \citet{2018Yang} focused on proton energies near the threshold of pion production up to 10 GeV in order to accurately prescribe the pion decay bump. They give a parametrization for the normalised pion energy distribution  $\tilde{f}(x,T_{\mathrm{p}})$ using the hadronic interaction model from the Geant4 Toolkit \citep{2003Agostinelli_Geant4,2006Allison_Geant4}, so that the differential cross section of pion production is given by
\begin{equation}
\frac{\mathrm{d}\sigma_{\pi}}{\mathrm{d}x}=\sigma_{\pi}\times \tilde{f}(x,T_{\mathrm{p}}),\label{eq:dsigma/dx Yang}
\end{equation}
where $x=T_{\pi}/T_{\pi,\mathrm{max}}$ is the ratio of the kinetic pion energy and $\sigma_{\pi}$ is the total cross section of pion production, including charged and neutral pions. There exist experimental data for the total cross section $\sigma_{\pi}$ below $T_{\mathrm{p}}\leq2\,\mathrm{GeV}$. Based on that, \citet{2014PhRvD..90l3014K} provide parametrizations in the energy range below $2\,\mathrm{GeV}$, where they include all neutral pion production channels, i.e., $\mathrm{pp}\rightarrow\mathrm{pp}\pi^{0}$, $\mathrm{pp} \rightarrow\mathrm{pp}2\pi^{0}$ as well as $\mathrm{p}\mathrm{p}\rightarrow\mathrm{p}\pi^{+}\pi^{0}$ and $\mathrm{p}\mathrm{p}\rightarrow D\pi^{+}\pi^{0}$. Above kinetic proton energies of $2\,\mathrm{GeV}$, the inelastic cross section is expressed in terms of the total inelastic cross section and an average pion multiplicity, that they fit separately, i.e.
\begin{equation}
\sigma_{\pi}=\sigma_{\mathrm{p}\mathrm{p},\mathrm{inel}}\left\langle n_{\pi}\right\rangle .\label{eq: sigma_pi}
\end{equation}
Here, $\left\langle n_{\pi}\right\rangle$ denotes the average pion multiplicity and the inelastic cross section is given by

\begin{align}
\sigma_{\mathrm{p}\mathrm{p},\mathrm{inel}}(T_{\mathrm{p}}) & =\left[30.7-0.96\log\left(\frac{T_{\mathrm{p}}}{T_{\mathrm{p},\mathrm{th}}}\right)+0.18\log^{2}\left(\frac{T_{\mathrm{p}}}{T_{\mathrm{p},\mathrm{th}}}\right)\right]\nonumber \\
 & \times\left[1-\left(\frac{T_{\mathrm{p}}}{T_{\mathrm{p},\mathrm{th}}}\right)^{1.9}\right]^{3}\,\mathrm{mbarn}, 
\label{eq:sigma_pp_inel Kafexhiu 2014}
\end{align}
where the threshold proton kinetic energy is $T_{\mathrm{p},\mathrm{th}}=2m_{\pi}c^{2}+m_{\pi}^{2}c^{4}/(2m_{\mathrm{p}}c^{2})\approx0.2797\,\mathrm{GeV}$. 
Additionally, \citet{2014PhRvD..90l3014K} provide a parametrization of the differential $\gamma$-ray cross section in the following form

\begin{equation}
\frac{\mathrm{d}\sigma_{\gamma}(T_{\mathrm{p}},E)}{\mathrm{d}E}=A_{\mathrm{max}}(T_{\mathrm{p}})F(T_{\mathrm{p}},E), \label{eq:dsigma_gamma/dE_gamma from Kafexhiu}
\end{equation}
where they fit $A_{\mathrm{max}}(T_{\mathrm{p}})=\max\left(\mathrm{d}\sigma_{\gamma}/\mathrm{d}E\right)$ separately from $F(T_{\mathrm{p}},E$) since the maximum value only depends on the proton kinetic energy $T_{\mathrm{p}}$. It is a function of the total $\pi^{0}$-production cross section $\sigma_{\pi}(E_{\mathrm{p}})$, for which they also provide their own fits. In the high-energy regime, they divide the cross section into the inelastic part and the pion multiplicity, see Eq.~(\ref{eq: sigma_pi}), and use Eq.~(\ref{eq:sigma_pp_inel Kafexhiu 2014}) for $\sigma_{\mathrm{pp, inel}}$.
This matches new experimental data by \citet{PhysRevD.86.010001} in the very high energy regime around $T_{\mathrm{p}}=10^{7}\,\mathrm{GeV}$ better than e.g., the one used by \citet{2006PhRvD..74c4018K}. They furthermore provide their own fit to the average pion multiplicity $\left\langle n_{\pi^{0}}\right\rangle $, that agrees well with the description used by \citet{2018Yang}, which refers to data from \citet{2001PAN....64.1841G}.

In our approach, we use the parametrization by \citet{2018Yang} for $T_{\mathrm{p}}<10\mathrm{\,GeV}$ and the model by \citet{2014PhRvD..90l3014K} at larger energies. We compare this to other models from the literature in App.~\ref{appendix:comparison1} and \ref{appendix:comparison2}. The relative deviation of the resulting total $\gamma$-ray luminosity, integrated from 0.1-100 GeV, of our model in comparison to the analytical approximation by \citet{2004A&A...413...17P} and the parametrization by \citet{2006PhRvD..74c4018K} is shown to be $\sim 10$ per cent, depending on the spectral index of the CR proton spectrum (see Fig.~\ref{fig:L_gam_comparison}).

So far, we have assumed that the ambient gas consists of protons only. The effect of relativistic protons interacting with nuclei heavier than hydrogen was studied by \citet{2018Yang}. At high energies these interactions can be described by a sequence of binary nucleon-nucleon collisions according to the Glauber's multiple scattering theory \citep{1955Glauber,1966FrancoGlauber,1970GlauberMatthiae}. However, there are two additional processes at lower, sub-relativistic energies, for which there exists no self-consistent theory. On the one hand, intra-nuclear collisions can lead to the production of pions below the kinematic threshold, which is called sub-threshold pion production. On the other hand, so called direct photons are emitted, probably due to neutron-proton-bremsstrahlung during the early stage of the nuclear interaction. The cross sections for these processes have been parametrized by \citet{2016Kafexhiu}. \citet{2018Yang} used these parametrizations to analyse the contribution from heavy nuclei to the $\gamma$-ray emission from hadronic interactions from Galactic CR protons with the interstellar gas and found a very similar spectral shape when including heavy nuclei in comparison to only considering pp-interactions, but found an overall increased emissivity by a nuclear enhancement factor of $a_{\mathrm{nucl}}=1.8$. Using the definition of the number density of hydrogen $n_{\mathrm{H}} = X_{\mathrm{H}}\,\rho/m_{\mathrm{p}}$, the helium density $n_{\mathrm{He}} = (1-X_{\mathrm{H}})/4 \times\rho/m_{\mathrm{p}}$ and the mass fraction of hydrogen $ X_{\mathrm{H}} = 0.76$, the number density of target nucleons in the ISM is given by $n_{\mathrm{N}} = n_{\mathrm{H}} + 4n_{\mathrm{He}}= \rho/m_{\mathrm{p}}$. Hence, if we use $n_{\mathrm{N}}$ as the target density for hadronic interactions, we only need another factor of $1.8\times X_{\mathrm{H}} \approx 1.37$ to account for the interactions of heavier nuclei, such as the sub-threshold pion production.

\subsection{Inverse Compton emission} \label{subsec:IC-emission}

The CR electron population also contributes to the non-thermal $\gamma$-ray emission. Inverse Compton (IC) scatterings transfer some of the CR electron energy to low-energy ambient photons, boosting them to very high energies up to the $\gamma$-ray regime. The typical energy gain for the up-scattered photon is given by $E_{\mathrm{max}}\approx4\gamma_{\mathrm{e}}^{2}E_{\mathrm{i}}/3$, where $\gamma_{\mathrm{e}}=\sqrt{p_{\mathrm{e}}^2+1}$ is the Lorentz factor of the electron colliding inelastically with an incident photon of energy $E_{\mathrm{i}}$. Following \citet{1968PhRv..167.1159J} and \citet{1970BlumenthalGould}, one can derive the emitted spectrum due to IC scattering of an electron population off of an incoming radiation field of photons, denoted by $n_{\mathrm{ph}}$. Using the general expression for the cross section of IC scattering described in the Klein-Nishina formalism, we can derive the IC emissivity resulting from a CR electron spectrum $N_{\mathrm{e}}(E_{\mathrm{e}})$ to obtain
\begin{align}
j_{\nu,\mathrm{IC}}&=2\upi h E r_{0}^{2}c \int\mathrm{d}E_{\mathrm{e}}\frac{f_{\mathrm{e}}(E_{\mathrm{e}})}{\gamma_{\mathrm{e}}^{2}}\nonumber\\ &\times\intop_{q_{\mathrm{min}}}^{1}\frac{\mathrm{d}q}{q}n_{\mathrm{ph}}[q(\gamma_{\mathrm{e}}, E)]\,f[q(\gamma_{\mathrm{e}}, E)], 
\label{eq: IC emission}
\end{align}
where $r_0=e^2/(m_{\mathrm{e}}c^2)$ is the classical electron radius, $e$ is the elementary charge, $m_{\mathrm{e}}$ the electron rest mass and 
\begin{equation}
f(q)=2q\ln q+(1+2q)(1-q)+\frac{1}{2}\frac{(\Gamma_{\mathrm{e}}q)^{2}}{1+\Gamma_{\mathrm{e}}q}(1-q). \label{eq: f(q)}
\end{equation}
Furthermore, the normalised quantities $\Gamma_{\mathrm{e}}$ and $q$ are given by $\Gamma_{\mathrm{e}}=4E_{\mathrm{i}}\gamma_{\mathrm{e}}/(m_{\mathrm{e}}c^{2})$and $ q(\gamma_{\mathrm{e}})=E^{*}m_{e}c^{2}/(4E_{\mathrm{i}}\gamma_{\mathrm{e}}(1-E^{*}))$, where $E^{*}=E/(\gamma_{\mathrm{e}} m_{\mathrm{e}}c^{2})$. The integration limits for $q$ follow from the kinematic limitations for $E^{*}$, i.e. $E_{\mathrm{i}}/(\gamma_{\mathrm{e}} m_{\mathrm{e}}c^{2})\leq E^{*}\leq\Gamma_{\mathrm{e}}/(1+\Gamma_{\mathrm{e}})$, where $\Gamma_{\mathrm{e}}=4E_{\mathrm{i}}\gamma_{\mathrm{e}}/(m_{\mathrm{e}}c^{2})$, so that $q_{\mathrm{min}}=\left[4\gamma_{\mathrm{e}}^{2}(1-E_{\mathrm{i}}/(\gamma_{\mathrm{e}} m_{\mathrm{e}}c^{2}))\right]^{-1}\leq q\leq1$.

The incident radiation field $n_{\mathrm{ph}}$ is assumed to consist of different components that can each be described by black body distributions of different temperatures $T_j$ and weighting factors $A_j$, such that
\begin{equation}
n_{\mathrm{ph}}(E)=\sum_{j}A_{j}\frac{E_{\mathrm{i}}^{2}}{\upi^{2}(\hbar c)^{3}\left(\exp(E/k_\rmn{B}T_{j})-1\right)},
\label{eq: radiation field}
\end{equation}
where $\hbar$ is the reduced Planck constant and $k_\rmn{B}$ is Boltzmann's constant.
For a given photon energy density $\varepsilon_{\mathrm{ph}}$, the black body spectrum is weighted according to
\begin{equation}
A_j=\varepsilon_{\mathrm{ph}}/(a_{\mathrm{rad}}T_j^{4}), 
\label{eq: photon energy density}
\end{equation}
where $a_{\mathrm{rad}}=8\upi^{5}k_{\mathrm{B}}^{4}/(15h^{3}c^{3})$ is the radiation constant.
To speed up the numerical integration of Eq.~(\ref{eq: IC emission}), we pre-evaluate the integral over $q$ for fixed momentum bins of the electron distribution, assuming two fixed black body temperatures, i.e.\  $T_{\mathrm{CMB}}=2.73\,\mathrm{K}$ and $T_{\mathrm{FIR}}=20\, \mathrm{K}$. The latter results from the assumption that the UV emission of young stars is absorbed by dust \citet{2000Calzetti} and re-emitted in the FIR. To obtain an estimate of the FIR flux in each cell of our simulation, we adopt the relation between SFR and FIR luminosity, $L_\rmn{FIR}$, obtained by \citet{1998Kennicutt}:
\begin{equation}
\frac{\dot{M}_{\star}}{\mathrm{M}_{\odot}\mathrm{\,yr^{-1}}} =a_{\rmn{SFR}}\,1.7\times10^{-10}\frac{L_{\mathrm{FIR}}}{L_{\odot}}.
\end{equation}
The parameter $a_{\rmn{SFR}}=0.79$ follows from adopting a \citet{2003Chabrier} initial mass function, see \citet{2010Crain}. The resulting photon energy density of a cell is then the sum of CMB and stellar radiation,i.e.\ $\varepsilon_{\mathrm{ph}}=\varepsilon_{\mathrm{CMB}}+\varepsilon_\star$, where the latter is derived by summing up the flux arriving at each cell from all actively star-forming cells at a distance $R_{i}$, i.e. $\varepsilon_\star=\sum_{i} L_{\mathrm{FIR}} / (4\upi R_{i}^{2}c)$. This flux is calculated using a tree code in order to speed up the calculation. If the considered cell is actively star forming, the distance is estimated from the cell's volume, $R_i=[3V_i/(4\upi)]^{1/3}$.

\subsection{Bremsstrahlung} \label{subsec:Bremsstrahlung}
The second process contributing to the non-thermal $\gamma$-ray emission of CR electrons is relativistic bremsstrahlung, which results from their acceleration in the field of charged nuclei. In the classical picture of the method of virtual quanta, i.e.\ the Weizs\"acker-Williams approach, it can be described as an IC scattering process in the rest frame of a relativistically moving electron. It sees the electrostatic field of the approaching nucleon with charge $Ze$ as a pulse of electromagnetic radiation, of which it inelastically scatters off and consequently, emits radiation while losing some of its energy. Therefore, in that picture one can again apply the IC formalism in the general Klein-Nishina regime, but replacing the incident radiation field with the virtual quanta from the approaching nucleus as seen from the electron's rest frame and  calculate the scattered photon field. The differential cross section for relativistic bremsstrahlung, $\mathrm{d}\sigma_{\mathrm{brems}}$, in the lab frame for an electron scattering off of a nucleon is thus given by $\mathrm{d}\sigma_{\mathrm{brems}}=\mathrm{d}\sigma_{\mathrm{IC}}\mathrm{d}N$. Inserting the Klein-Nishina cross section and the expression for the differential number of virtual quanta $\mathrm{d}N$, one has to integrate over the impact parameter \citep[see][]{1970BlumenthalGould}. Eventually, a Lorentz transformation back to the lab frame yields
\begin{align}
\mathrm{d}\sigma_{\mathrm{brems}}=4\alpha r_{0}^{2}Z^{2}\frac{\mathrm{d}\omega}{\omega} \left[\frac{4}{3}\left(1-\frac{\hbar\omega}{E_{\mathrm{e,in}}}\right)+\left(\frac{\hbar\omega}{E_{\mathrm{e,in}}}\right)^{2}\right]
\nonumber \\
\times \ln\left(\frac{2E_{\mathrm{e,in}}E_{\mathrm{e,out}}}{m_{\mathrm{e}}c^{2}\hbar\omega}\right),
\label{eq: Brems. dsigma final}
\end{align}
where $\alpha$ is the fine structure constant, $\omega$ is the angular frequency of the emitted photon and $E_{\mathrm{e,in}}$ denotes the energy of the incoming electron, whereas $E_{\mathrm{e,out}}$ is the energy of the outgoing electron after scattering, i.e., $E_{\mathrm{e,out}}=E_{\mathrm{e,in}}-\hbar \omega$.
Since the argument of the logarithm is typically $\gg1$, this is consistent with the cross section derived by \citet{1934BetheHeitler} in the Born approximation for non-screened (i.e.\ fully ionized) ions and for the case of highly relativistic electrons, i.e.
\begin{align}
\mathrm{d}\sigma_{\mathrm{brems}}=4\alpha r_{0}^{2}Z^{2}\frac{\mathrm{d}\omega}{\omega}\frac{1}{E_{\mathrm{e,in}}^{2}}\left(E_{\mathrm{e,in}}^{2}+E_{\mathrm{e,out}}^{2}-\frac{2}{3}E_{\mathrm{e,in}}E_{\mathrm{e,out}}\right)
\nonumber \\
\times \left(\ln\frac{2E_{\mathrm{e,in}}E_{\mathrm{e,out}}}{m_{e}c^{2}\hbar\omega}-\frac{1}{2}\right).\label{eq: sigma brems. Born approx. extr. rel.}
\end{align}
As pointed out by \citet{1997Haug}, if we consider mildly and highly relativistic electrons, one can combine the non-relativistic cross section obtained by \citet{1954Heitler} with the extreme relativistic case that is expanded up to 6th orders in initial and final electron momenta. 
Equation\,(\ref{eq: sigma brems. Born approx. extr. rel.}) has a relative deviation of order  $<10^{-3}$ for frequencies above $0.1\,\mathrm{GeV}$, when comparing it to the extension of the semi-relativistic description in \citet{1997Haug}, that also includes an correction factor that accounts for the distortion of the electron wave function in the Coulomb field of the nucleus. Hence, we adopt this approximation in the following.

In addition to electron-ion bremsstrahlung, we also have to take into account electron-electron bremsstrahlung. While at low incident electron energies the quadrupole emission from the electron-electron interaction can be neglected in comparison to the electron-nucleus dipole emission, it can make a significant contribution for higher energy electrons and emitted photons \citep{1975bHaug}.
The exact expression for the electron-electron bremsstrahlung cross section $\mathrm{d}\sigma_{ee}$ was first derived by \citet{1975aHaug}, for which \citet{SovietPhys.24.760} provided a good approximation for ultra-relativistic electrons. It is given in terms of the normalised energy $\epsilon=E/(m_{\mathrm{e}}c^{2})$, where $E$ denotes the energy of the scattered photon, by
\begin{equation}
\mathrm{d}\sigma_{ee}=(\sigma_{1}+\sigma_{2})A(\epsilon,\gamma_{\mathrm{e}})\mathrm{d}\epsilon, \label{eq:d_sigma_electron-electron-bremsstrahlung}
\end{equation}
where
\begin{equation}
\sigma_{1}=\frac{4r_{0}^{2}\alpha}{\epsilon}\left[\frac{4}{3}(1-\frac{\epsilon}{\gamma_{\mathrm{e}}})+\left(\frac{\epsilon}{\gamma_{\mathrm{e}}}\right)\right]\left[\ln\frac{2\gamma_{\mathrm{e}}(\gamma_{\mathrm{e}}-\epsilon)}{\epsilon}-\frac{1}{2}\right]\label{eq: ee-bremsstrahlung sigma_1}
\end{equation}
and
\begin{align}
\sigma_{2} = \frac{r_0^{2}\alpha}{3\epsilon} \begin{cases}
\begin{array}{c}
16(1-\epsilon+\epsilon^{2})\ln\frac{\dps\gamma_{\mathrm{e}}}{\dps\epsilon} -\frac{\dps1}{\dps\epsilon^{2}}+\frac{\dps3}{\dps\epsilon}-4\\
 +4\epsilon - 8\epsilon^{2}-2(1-2\epsilon)\ln\left(1-2\epsilon\right)\\
 \left(\frac{\dps1}{\dps4\epsilon^{3}}-\frac{\dps1}{\dps2\epsilon^{2}}+ \frac{\dps3}{\dps\epsilon} - 2 + 4\epsilon\right),\\
\\
\frac{\dps2}{\dps\epsilon}\left[\left(4-\frac{\dps1}{\dps\epsilon}+\frac{\dps1}{\dps4\epsilon^{2}}\right)\ln2\gamma_{\mathrm{e}}\right.\\
\left. -2+\frac{\dps2}{\dps\epsilon}-\frac{\dps5}{\dps8\epsilon^{2}} \right],
\end{array} & \begin{array}{c}
\epsilon\leq\frac{\dps1}{\dps2}\\
\\
\\
\epsilon>\frac{\dps1}{\dps2}.
\end{array}\end{cases} \label{eq:ee-bremsstrahlung sigma_2}
\end{align}
The factor $A(\epsilon,\gamma_{\mathrm{e}})$ is a middly-relativistic correction factor that was introduced by \citet{1999ApJ...513..311B} and reads 
\[
A(\epsilon,\gamma_{\mathrm{e}})=1-\frac{8}{3}\frac{\left(\gamma_{\mathrm{e}}-1\right)^{1/5}}{\gamma_{\mathrm{e}}+1}\left(\frac{\epsilon}{\gamma_{\mathrm{e}}}\right)^{1/3}.
\]
According to them, the highly relativistic approximation combined with this factor yields an accuracy within 10 per cent in comparison to the exact expression by \citet{1975aHaug}.
We finally arrive at the bremsstrahlung emissivity resulting from a CR electron population as
\begin{equation}
j_{\nu,\mathrm{brems}}=j_{\nu,\mathrm{ep}}+j_{\nu,\mathrm{ee}},
\end{equation}
where we take into account the contribution of electron-proton and electron-electron bremsstrahlung, which are calculated by
\begin{equation}
j_{\nu,\mathrm{ep/ee}}= c\,n_{\mathrm{p/e}}\nu h^2\int \mathrm{d}p_{\mathrm{e}} f_{\mathrm{e}}(p_{\mathrm{e}})\frac{\mathrm{d}\sigma_{\mathrm{ep/ee}} (p_{\mathrm{e}},\nu)}{\mathrm{d}\epsilon}.
\label{eq:j_brems}
\end{equation}

\section{Comparison to hadronic interaction models}
\label{sec: comparison to other models}

\begin{figure*}
\begin{centering}
\includegraphics[scale=1]{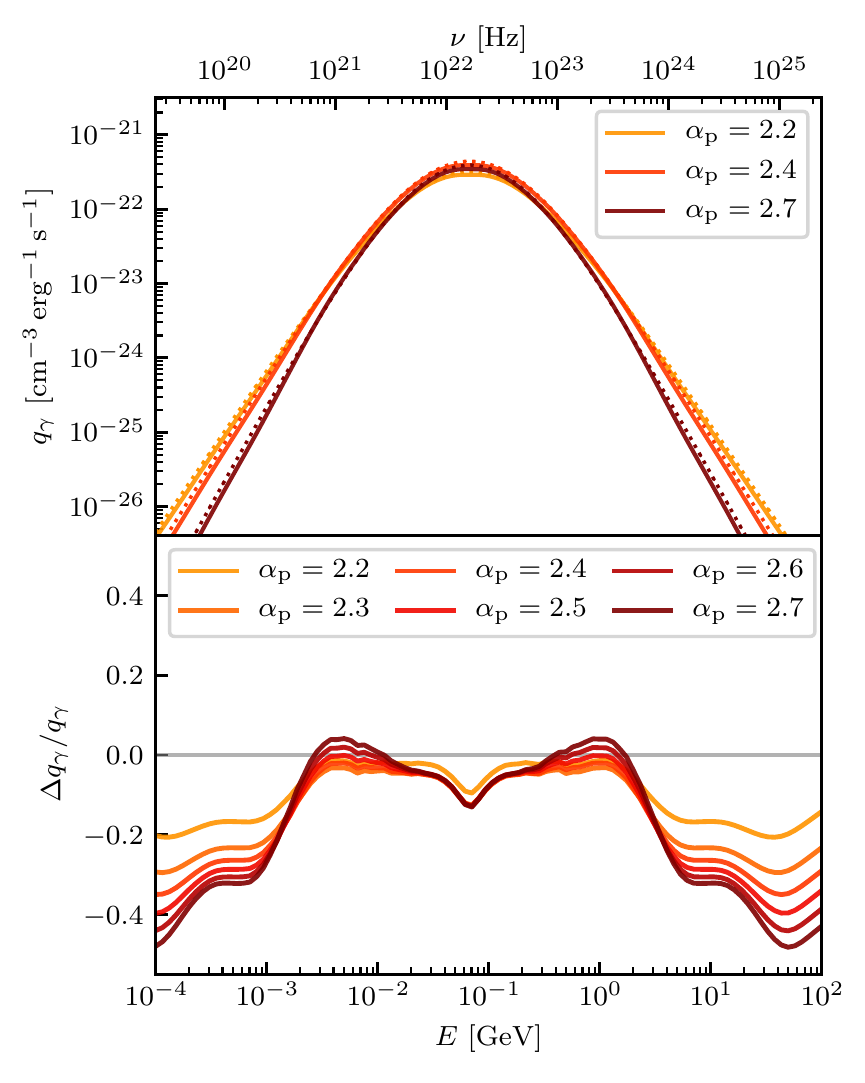}
\includegraphics[scale=1]{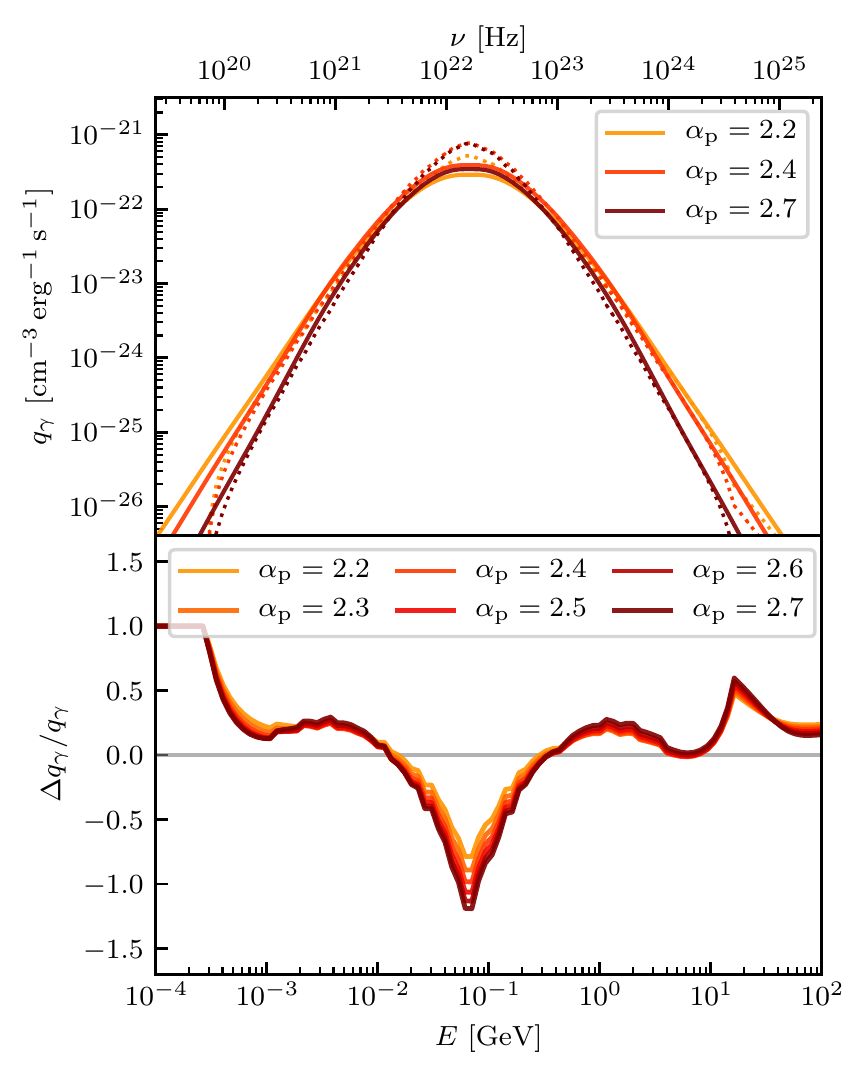}
\par\end{centering}
\caption{The $\gamma$-ray emissivity for a power-law distribution of protons with different spectral indices, normalised to $\varepsilon_{\mathrm{CR}}=1\,\mathrm{eV\,cm^{-3}}$ and $n_{\mathrm{N}}=1\,\mathrm{cm^{-3}}$, resulting from our approach (solid lines) and from the analytical approximation by \citet{2004A&A...413...17P}  (dotted lines, left panel), and the model by \citet{2006PhRvD..74c4018K} (dotted lines, right panel). The lower panels shows their relative differences, respectively. Note the different scales in both panels.} \label{fig:j_gam_comparison}
\end{figure*}
\begin{figure}
\begin{centering}
\includegraphics[scale=1]{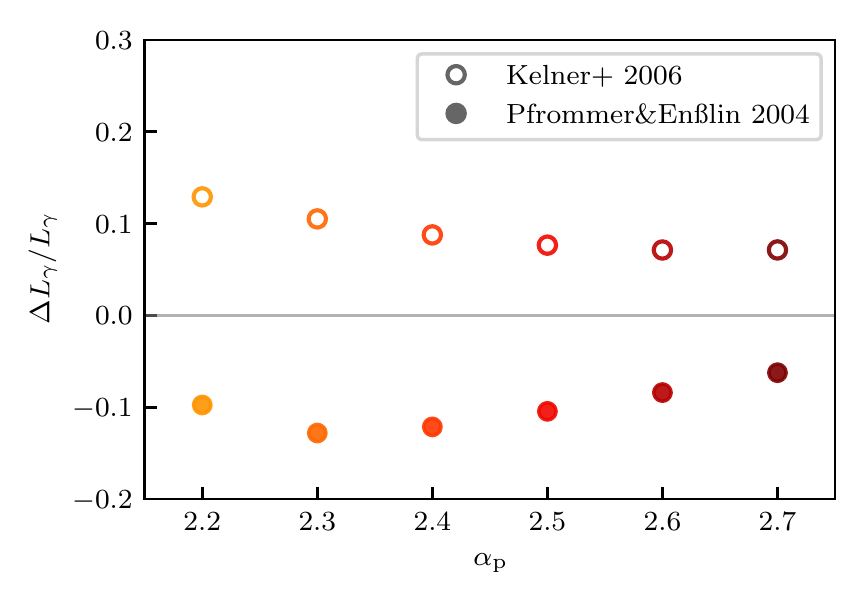}
\par\end{centering}
\caption{Relative deviation of the total luminosity $L_{\gamma}$ between 0.1-100 GeV calculated from our model in comparison to \citet{2006PhRvD..74c4018K} (open symbols) and \citet{2004A&A...413...17P} (filled symbols) for different values of  the proton spectral index $\alpha_{\mathrm{p}}$ (see legend in Fig.~\ref{fig:j_gam_comparison}).} 
\label{fig:L_gam_comparison}
\end{figure}

\subsection{Analytical approximation by \citet{2004A&A...413...17P} \label{appendix:comparison1} }
\citet{2004A&A...413...17P} derived an analytical expression for the $\gamma$-ray source function. They aimed for connecting the high energy limits for the $\gamma$-ray source function to the detailed physics near the threshold of neutral pion production that have been modeled with the COSMOCR code \citep{2001CoPhC.141...17M}, that is based on the isobaric model and also takes into account the contribution of kaon decay modes to the neutral pion production. The resulting analytical formula assumes a power-lar CR momentum distribution that extends below the kinematic threshold for the pp reaction and reads \citep{2004A&A...413...17P}
\begin{align}
\begin{split}
q_{\gamma}\simeq \sigma_{\mathrm{p}\mathrm{p}}c n_{\mathrm{N}}\xi^{2-a_{\gamma}} C_{\mathrm{p}}\frac{4}{3a_{\gamma}} \left(\frac{m_{\pi}}{m_{\mathrm{p}}}\right)^{-a_{\gamma}} \\ 
\times\left[ \left(\frac{2E}{m_{\pi}c^{2}}\right)^{\delta_{\gamma}} + \left(\frac{2E}{m_{\pi}c^{2}}\right)^{-\delta_{\gamma}} \right],
\end{split}
\label{eq: gamma-ray source fct. analytical Pfrommer}
\end{align}
where $n_{\mathrm{N}}$ denotes the target nucleon density, the pion multiplicity is assumed to be constant, $\xi=2$, and $C_{\mathrm{p}}$ is the normalization of the proton momentum distribution at momentum $m_{\mathrm{p}}c$. The asymptotic slope of the $\gamma$-ray spectrum $a_{\gamma}$ equals the spectral index of the proton population in the scaling model, that has also been adopted by \citet{1986ApJ...307...47D}. Furthermore, the parameter $\delta_{\gamma}$ and the total effective cross section $\sigma_{\mathrm{pp}}$ have been modeled by \citet{2004A&A...413...17P} as
\begin{align}
\delta_{\gamma}=0.14a_{\gamma}^{-1.6}+0.44\label{eq: delta_gamma Pfrommer}
\end{align}
and
\begin{equation}
\sigma_{\mathrm{p}\mathrm{p}}=32\times(0.96+\exp(4.4-2.4a_{\gamma})\,\mathrm{mb.}\label{eq: sigma_pp Pfrommer}
\end{equation}
The analytical approximation by \citet{2004A&A...413...17P} from Eq.~(\ref{eq: gamma-ray source fct. analytical Pfrommer}) slightly over-predicts the resulting emission near the threshold of pion production, i.e. at $m_\pi c^{2}/2\approx67.5\,\mathrm{MeV}$(see Fig.~\ref{fig:j_gam_comparison}), as well as at very high and very low $\gamma$-ray energies. However, the total $\gamma$-ray luminosity between 0.1-100\,GeV is accurate to 10 percent in comparison to our approach (see Fig.~\ref{fig:L_gam_comparison}).

\subsection{Parametrization by \citet{2006PhRvD..74c4018K} \label{appendix:comparison2} }
Frequently used analytical expressions for the energy spectra of secondary particles in pp-collisions like pions, electrons, neutrinos and $\gamma$ rays are provided by \citet{2006PhRvD..74c4018K}. They focus on the high energy regime, where $T_{\mathrm{p}}>100\mathrm{\,GeV}$, and prescribe the production of $\gamma$ rays in terms of the number of created photons in the interval $(x,x+\mathrm{d}x)$ per collision, denoted by $F_{\gamma}(x,E_{\mathrm{p}})$, with $x=E_{\gamma}/E_{\mathrm{p}}$. 
By convolving $F_{\gamma}$ with the proton energy distribution $N_{\mathrm{p}}(\gamma_{\mathrm{p}})$ and the inelastic cross section of pp-interactions, one obtains the $\gamma$-ray production rate in the energy interval $(E_{\gamma},E_{\gamma}+dE_{\gamma})$ via
\begin{equation}
q_{\gamma}(E_{\gamma})= cn_{\mathrm{H}}\intop_{E_{\gamma}}^{\infty}\sigma_{\mathrm{inel}}(E_{\mathrm{p}}) f_{\mathrm{p}}(E_{\mathrm{p}})F_{\gamma}\left(\frac{E_{\gamma}}{E_{\mathrm{p}}},E_{\mathrm{p}}\right)\frac{\mathrm{d}E_{\mathrm{p}}}{E_{\mathrm{p}}}.\label{eq:gamma-ray-production-rate Kelner}
\end{equation}
Fitting the numerical data with the SIBYLL code \citep{1994Fletcher_SIBYLL}, they obtain the inelastic part of the total cross section of pp-interactions, $\sigma_{\mathrm{inel}}(E_{\mathrm{p}})/\mathrm{mb}=34.3+1.88L+0.25L^{2}$. In order to better match the experimental data in the low energy regime, they multiply this expression by a factor of $(1-\left(E_{\mathrm{th}}/E_{\mathrm{p}}\right)^{4})^{2}$. 
For low proton kinetic energies ($T_{\mathrm{p}}<100\mathrm{GeV}$), where their parametrization for $F_{\gamma}(x,E_{\mathrm{p}})$ is not valid, they suggest a $\delta$-functional approach for the production rate of pions. Hence, in this approximation, the production rate of pions is given by 
\begin{align}
\tilde{F}_{\pi}(E_{\pi},E_{\mathrm{p}})=\tilde{n}\delta\left(E_{\pi}-\frac{\kappa}{\tilde{n}}E_{\mathrm{kin}}\right),
\end{align}
where $\tilde{n}=\int\tilde{F_{\pi}}dE_{\pi}$ denotes the number of produced pions and $\kappa$ is the fraction of kinetic energy transferred to $\gamma$ rays by a proton with energy $E_{\mathrm{p}}$. From this, they obtain the pion source function via
\begin{equation}
q_{\pi}(E_{\pi})=\tilde{n}\frac{cn_{\mathrm{H}}}{K_{\pi}}\sigma_{\mathrm{inel}}\left(m_{\mathrm{p}}c^{2}+\frac{E_{\pi}}{K_{\pi}}\right)
f_{\mathrm{p}}\left(m_{\mathrm{p}}c^{2}+\frac{\dps E_{\pi}}{\dps K_{\pi}}\right).
\end{equation}
The parameter $K_{\pi}=\kappa/\tilde{n}$ prescribes the mean fraction of proton energy that is transferred to the produced neutral pion and is assumed to be $K_{\pi}=0.17$, which agrees well with numerical Monte Carlo simulations \citep{1997Mori}, as demonstrated in \citet{2000Aharonian}.
From this expression we can calculate the resulting $\gamma$-ray spectrum with
\begin{equation}
q_{\gamma}(E) = 2\intop_{E_{\pi,\mathrm{min}}}^{E_{\pi,\mathrm{max}}}\mathrm{d}E_{\pi}\frac{q_{\pi}(E_{\pi})}{\sqrt{E_{\pi}^{2}-(m_{\pi}c^{2})^{2}}},\label{eq:q_gamma from q_pion}
\end{equation}

In comparison to our approach, which is shown by the solid lines in Fig.~\ref{fig:j_gam_comparison}, the $\delta$-function approximation by \citet{2006PhRvD..74c4018K} for low proton energies can be recognized by a sharp peak around the pion-decay bump and hence overproduces the $\gamma$-ray emission at these energies in comparison to our approach by more than 100 per cent. The transition from the $\delta$-approximation to their parametrization at higher proton energies > 100 GeV is visible in the $\gamma$-ray source function at energies around 10 GeV. The effect on the total $\gamma$-ray luminosity is depicted in Fig.~\ref{fig:L_gam_comparison}. It shows that our approach yields an about 10\% higher $\gamma$-ray luminosity in comparison to the approach by \citet{2006PhRvD..74c4018K}.

\bsp

\label{lastpage}
\end{document}